\documentclass[11pt,a4paper]{article}
\pdfoutput=1
\usepackage{jheppub}
\usepackage{amsmath,amssymb}
\usepackage{graphicx} 
\usepackage{comment}
\allowdisplaybreaks[1]

\title{Triangulation of 2-loop MHV Amplituhedron from Sign Flips}

\author{Ryota Kojima}
\affiliation{Department of Particle and Nuclear Physics\\SOKENDAI (The Graduate University for Advanced Studies)\\ Tsukuba, Ibaraki, 305-0801, Japan\\
KEK Theory Center, Tsukuba, Ibaraki, 305-0801, Japan}

\emailAdd{ryota@post.kek.jp}

\abstract{
In this paper, we consider the triangulation of the 2-loop MHV amplituhedron from "sign flip" definition. Using the isomorphism between the $m=2, k=2$ tree amplituhedron and the 1-loop MHV physical amplituhedron, we found the direct triangulation of the 2-loop MHV amplituhedron from sign flips. This triangulation is different from the BCFW triangulation because of the structure of the cells. And we also found a formula of the canonical form of the $n$-point 2-loop MHV amplituhedron. This formula looks like a 2-loop version of the Kermit representation of the 1-loop MHV amplitude.  We checked that the sum of these cells is consistent with the BCFW up to at least 22-pt numerically.}

\preprint{\begin{flushright}
KEK-TH-2088
\end{flushright}}

\begin{document}
\maketitle
\newpage


\section{The Amplituhedron and Triangulation}
Recent years an unexpected connection between the scattering amplitudes and new geometric structure have been revealed \cite{spuriouspole,notepolytope,positivegrassmannian,positivegeometry}. In planar $\mathcal{N}=4$ SYM theory, the amplituhedron gives a definition of the scattering amplitude in a purely geometric way \cite{amplituhedron,intoamplituhedron} and it has been explored from a variety of perspectives in the past few years (see e.g.\cite{deepamplituhedron,
Karp:2017ouj, Karp:2016uax, Heslop:2018nht, Rao:2018uta, Galashin:2018fri, An:2017tbf,Rao:2017fqc,Galloni:2016iuj,Franco:2014csa,Bai:2014cna}). The amplituhedron is a geometric object that conjectured to give the scattering amplitude in planar $\mathcal{N}=4$ SYM theory as its form related to the volume: the canonical form. The canonical form is defined that it has logarithmic singularities on all the boundaries of the amplituhedron \cite{positivegeometry}.  \par
There are two definitions of the amplituhedron, a generalization of the interior of plane polygons into the Grassmannian \cite{amplituhedron} and a topological definition "sign flip definition" \cite{windingamplituhedron}. To see the first definition, we consider a convex $n$-polygon in projective space. The vertices of this polygon can be represented as 3-vectors $Z^I_a$ for $I=1,2,3$ and $a=1,2,\cdots,n$. The convexity means that all ordered minors $\langle Z_aZ_bZ_c \rangle>0$ for $a<b<c$ are positive. Then the interior of this polygon can be thought as the set of points $Y^I$ that $Y^I=c_aZ_a^I$ with $c_a>0$. The tree amplituhedron $\mathcal{A}_{m,k,n}$ can be obtained as the generalization of this structure into the Grassmannian
\begin{equation}
Y^I_\alpha =C_{\alpha a}Z^I_a,\ \ \ \text{for}\ \ \ I=1,\cdots,k+m,\ \ a=1,\cdots,n
\end{equation}
where $C$ is the positive Grassmannian $G_+(k,n)$ and $Z$ is the positive external data 
\begin{equation}
\langle Z_{a_1},\cdots,Z_{a_{k+m}} \rangle>0\ \text{for}\ \ a_1,\cdots<a_{k+m}.
\end{equation}
We can generalize this tree amplituhedron into the loop amplituhedron. We fix $m=4$, then $Y$ is the $k$-plane in k+4 dimensional space. Let's consider $L$ 2-dimensional planes $\mathcal{L}_{(i)}$ in 4-dimensional space complement of $Y$. These $\mathcal{L}$ are  the different linear combination of the external data $\mathcal{Z}$ 
\begin{equation}
\mathcal{L}^I_{(i)\alpha}=D_{a\alpha(i)}Z^I_a
\end{equation}
where D is the positive Grassmannian $G_+(2,n)$. Then the full amplituhedron $\mathcal{A}_{n,k,l}$ is the space of all $Y$, $\mathcal{L}_{(i)}$ of the form
\begin{equation}
Y^I_\alpha =C_{\alpha a}Z^I_a,\ \ \ \ \ \mathcal{L}^I_{(i)\alpha}=D_{a\alpha(i)}Z^I_a
\end{equation}
where all ordered minors of the matrix
\begin{equation}
\left(
    \begin{array}{ccccc}
     D_{(i_1)}\\
     D_{(i_2)}\\
     \vdots\\
     D_{(i_l)}\\
     C\\
    \end{array}
  \right)
\end{equation}
are positive. \par
Next, we consider the topological definition. Polytopes can be defined in two different ways. The first is called "vertex-centered": the polytopes is defined as the convex full of the points $Z_a^I$. This related to the $"Y^I=C_aZ_a^I"$ description of the amplituhedron. Second is a "face-centered" description of the polytope: we can obtain the polytope by the collection of inequalities associated with the facet of the polytope. This picture is generalized to the amplituhedron \cite{windingamplituhedron}. In addition to the boundary inequalities, sign flip characterization is needed in the case of the amplituhedron. The face-centered definition of the $m=2$ tree amplituhedron is 
\begin{center}
$Y$ is in the $m=2$ amplituhedron iff\\
$\langle Y i i+1 \rangle>0$ and the sequence $\{\langle Y12 \rangle, \cdots,\langle Y1n \rangle \}$ has precisely $k$ sign flips. 
\end{center}
We can define $m=4$ amplituhedron similarly. The sign flip definition of the loop amplituhedron is
\begin{eqnarray}
\label{eq:defloop}
\langle (YAB)_\gamma ii+1 \rangle >0,\ \langle Yii+1jj+1 \rangle >0&\nonumber\\
\{(\langle YAB)_\gamma12 \rangle, \cdots,\langle (YAB)_\gamma1n \rangle \} \ \text{has} \ k+2 \ \ \text{sign flips}&\\
\langle Y1234 \rangle >0,\cdots,\langle Y123n \rangle \ \text{ has}\  k\  \text{sign flips}\nonumber
\end{eqnarray}
where $\gamma$ is the number of loops. \par
To obtain the scattering amplitude from this geometry, we need to consider how to obtain the canonical form from this geometry. There are several ways: first is to obtain the canonical form directly from the geometry \cite{positiveamplituhedron,towardvolume,jkresidue}. Another way is to triangulate the amplituhedron into a more simple one that its form is trivial. The triangulation is related to the recursion relation of scattering amplitude. \par
Note that the $Y=C\cdot Z$ description of the amplituhedron is highly redundant. The space of $C_{\alpha a}$ is $k(n-k)$ dimensional and it is always larger than the dimension of the tree amplituhedron $k\times m$. Non-redundant map into $Y$ can only come from the $k\times m$ dimensional cells of $C\in G_+(k, n)$. It is difficult to obtain the canonical form of this highly redundant space directly, however, we can triangulate the amplituhedron into the non-redundant cells. It is straightforward to obtain the canonical form of this non-redundant cells; for example, the case of a triangle in the projective plane or the $n$-simplex in the $n$-dimensional projective space \cite{amplituhedron}.  Then we can obtain the canonical form of this non-redundant cell and the full form is obtained from the sum of these forms. \par
One of the important example of the triangulation of the amplituhedron is the BCFW triangulation. In planar $\mathcal{N}=4$ SYM, all loop order amplitudes can be obtained from the BCFW recursion relation \cite{loopbcfw} and there is a triangulation of the amplituhedron related to this BCFW recursion relation \cite{positivegrassmannian,amplituhedron,Bai:2014cna}. Another example is the direct triangulation. In the simple case, we can triangulate the amplituhedron directly from the geometry \cite{notepolytope}. In the case of the polygon: $\mathcal{A}(k=1,m=2,n)$, the dimension of non-redundant cells is $2$ and these cells are the triangle in the projective plane. Then the triangulation of the $\mathcal{A}(1,2,n)$ is straightforward from the geometry. Another example is the 1-loop MHV amplituhedron $\mathcal{A}(l=1,k=0,n)$. One of the triangulation of this case is known: Kermit representation
\begin{equation}
M^{\text{1-loop}}_{\text{MHV}}=\sum_{i<j}[1ii+1,1jj+1]
\end{equation}
where
\begin{equation}
[1ii+1,1jj+1]=\frac{\langle AB(1ii+1)\cap(ijj+1)\rangle^2}{\langle AB1i \rangle\langle AB1i+1 \rangle\langle ABii+1 \rangle\langle AB1j \rangle\langle AB1j+1 \rangle\langle ABjj+1 \rangle}.
\end{equation}
The dimension of each cell is $4$, then these cells are non-redundant cells. This triangulation is obtained directly from the geometry \cite{notepolytope}. However, the more general case, it is difficult to obtain the triangulation directly from the geometry.  To see this, we consider the 2-loop n-point MHV amplituhedron. In the $Y=C\cdot Z$ description we have many positivity conditions that 
\begin{equation}
\text{All ordered minor of}
\left(
    \begin{array}{ccccc}
     D_{(1)}\\
     D_{(2)}
    \end{array}
  \right)\text{are positive.}
\end{equation}
where $D_{(i)}\in G_+(2,n)$.  Because of these positivity conditions, it is difficult to obtain the triangulation for general $n$-point 2-loop amplitude directly from the geometry in this  $Y=C\cdot Z$ description. Of course, we can triangulate this from the BCFW approach and obtain the canonical form. For example, in the case of the 4-point, it begins with the $k=2$, $n=8$ tree amplitude and after taking two forward limits, we can obtain the 2-loop MHV amplitude. Then the question is that is it possible to obtain the triangulation of the 2-loop MHV amplituhedron directly from the geometry, without having to refer to any tree amplitude or the BCFW recursion relation? \par
 In this paper, we consider the triangulation of the $n$-point 2-loop MHV amplituhedron. we will give the triangulation of this amplituhedron from the sign flip definition and the full form of this amplituhedron as the sum of these cells. This triangulation is obtained directly from the geometry, without having to refer to any tree amplitude or the BCFW recursion relation. Then we will compare with the BCFW, local representation and our results.
\section{BCFW and Double Pentagon Expansions}
First we review briefly the BCFW recursion relation and double pentagon representation of the $n$-point 2-loop MHV amplitude. The BCFW recursion relation for all loop amplitudes in planar $\mathcal{N}=4$ SYM \cite{loopbcfw} is
\begin{eqnarray}
M_{n,k,l}(1,\cdots,n)&=&M_{n-1,k,l}(1,\cdots,n-1)\nonumber\\
&+&\sum_{n_L,k_L,l_L;j}[j\ j+1\ n-1\ n\ 1]M^R_{n_R,k_R,l_R}(1,\cdots,j,I_j)\times M^L_{n_L,k_L,l_L}(I_j,j+1,\cdots,\hat{n}_j)\nonumber\\
&+&\int_{\text{GL(2)}}[AB\ n-1\ n\ 1]\times M_{n+2,k+1,l-1}(1,\cdots,\hat{n}_{AB},\hat{A},B)
\end{eqnarray}
where $n_L+n_R=n,\ k_L+k_R=k,\ l_L+l_R=l$ and
\begin{eqnarray}
&&[a,b,c,d,e]=\frac{\delta^{0|4}(\eta_a\langle bcde \rangle+\text{cyclic})}{\langle abcd \rangle\langle bcde \rangle\langle cdea \rangle\langle deab \rangle\langle eabc \rangle}\nonumber\\
&&\hat{n}_j=(n-1\ n)\cap(j\ j+1\ 1),\ \ \ \ \ I_j=(j\ j+1)\cap(n-1\ n\ 1),\nonumber\\
&&\hat{n}_{AB}=(n-1\ n)\cap(A\ B\ 1),\ \ \ \ \ \hat{A}=(A\ B)\cap(n-1\ n\ 1).
\end{eqnarray}
Explicit BCFW representation of the 2-loop amplitude is written in \cite{all2loop}. This BCFW representation has spurious poles in each term. \par
The another representation is the local representation. Each term of the local representation has no spurious pole. The local representation of the 2-loop MHV amplitudes is obtained from the sum of the double pentagon diagrams \cite{localintegral}. The double pentagon diagram is
\begin{eqnarray}
  \raisebox{-1.25cm}{\includegraphics[width=4cm]{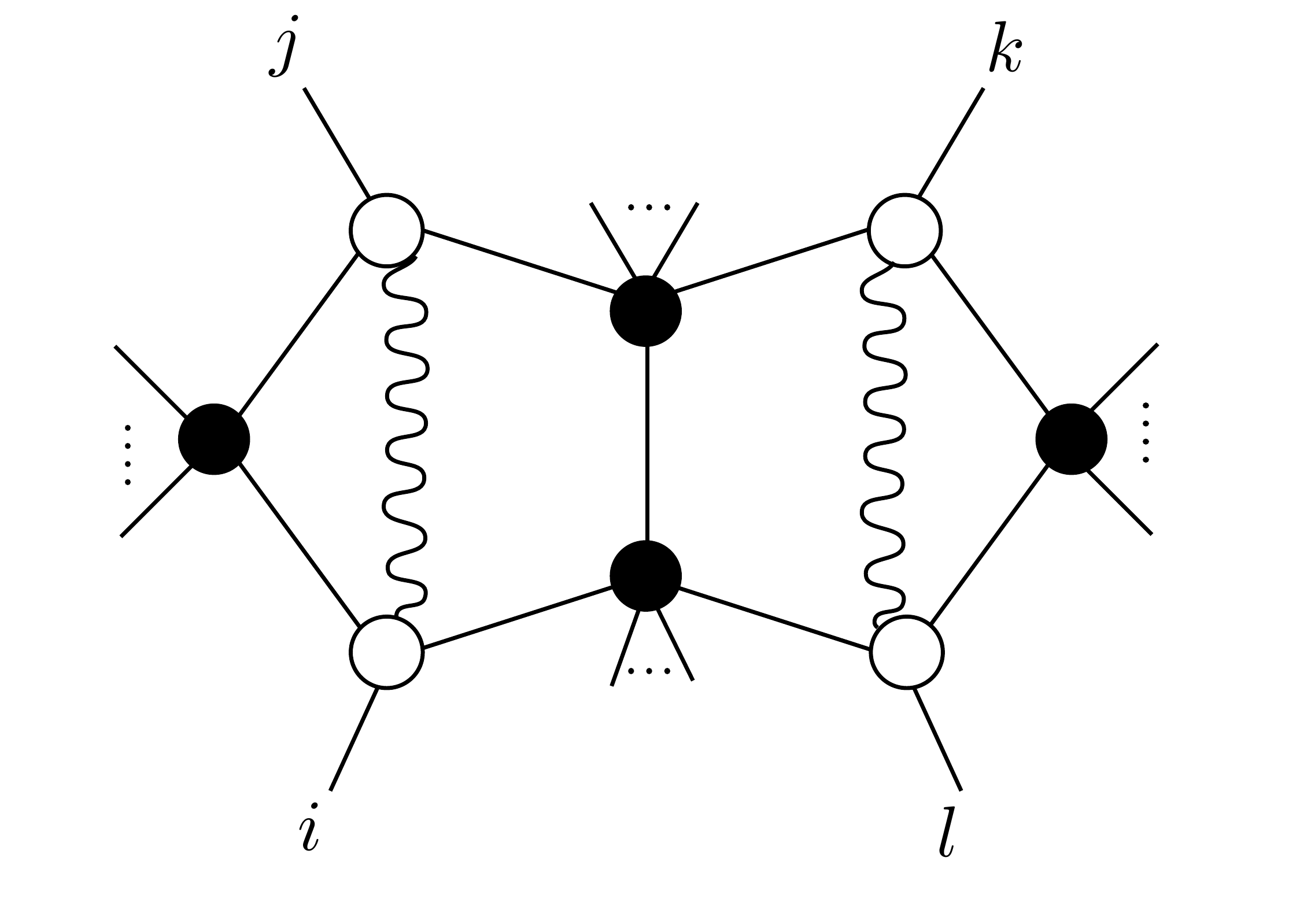}}&=&\LARGE{\substack{\frac{\langle AB(i-1ii+1)\cap(j-1jj+1)\rangle \langle ijkl \rangle}{\langle AB i-1i \rangle\langle ABii+1 \rangle \langle ABj-1j\rangle\langle ABjj+1 \rangle \langle ABCD \rangle}\nonumber\\\times\frac{\langle CD(k-1kk+1)\cap(l-1ll+1)\rangle }{\langle CD k-1k \rangle\langle CDkk+1 \rangle \langle CDl-1l\rangle\langle CDll+1 \rangle}}}\\
  &=&Q_{ijkl}.
\end{eqnarray}
This diagram has no spurious pole and the full amplitude is the cyclic sum over this diagram. For example, the local representation of the 2-loop MHV 4-point amplitude is
\begin{eqnarray}
\label{eq:local4pt}
\mathcal{A}^{2-\text{loop},\ 4-\text{pt}}_{\text{MHV}}
&=&Q_{1234}+Q_{2341}+Q_{3412}+Q_{4123}\nonumber\\
&=&\frac{\langle 1234 \rangle^3 }{\langle AB12 \rangle \langle AB14 \rangle\langle AB23 \rangle \langle ABCD \rangle \langle CD14 \rangle\langle CD23 \rangle \langle CD34 \rangle}\nonumber\\&+&\frac{\langle 1234 \rangle^3 }{\langle AB12 \rangle \langle AB14 \rangle\langle AB34 \rangle\langle ABCD \rangle\langle CD12 \rangle\langle CD23 \rangle \langle CD34 \rangle}\\&+&\frac{\langle 1234 \rangle^3 }{\langle AB12 \rangle\langle AB23 \rangle \langle AB34 \rangle\langle ABCD \rangle\langle CD12 \rangle \langle CD14 \rangle \langle CD34 \rangle}\nonumber\\&+&\frac{\langle 1234 \rangle^3 }{ \langle AB14 \rangle\langle AB23 \rangle\langle AB34 \rangle\langle ABCD \rangle\langle CD12 \rangle \langle CD14 \rangle\langle CD23 \rangle }.\nonumber
\end{eqnarray}
The 5-point case is
\begin{eqnarray}
\label{eq:local5pt}
\mathcal{A}^{2-\text{loop},\ 5-\text{pt}}_{\text{MHV}}&=&(1,2,3,4)+(2,3,4,5)+(3,4,5,1)+(4,5,1,2)
+(5,1,2,3)
\end{eqnarray}
where $(i,j,k,l)=Q_{ijkl}+Q_{jkli}+Q_{klij}+Q_{lijk}$. The general $n$-point amplitude is
\begin{eqnarray}
\label{eq:localnpt}
\mathcal{A}^{2-\text{loop}}_{\text{MHV}}=\sum_{i<j<k<l<i} \raisebox{-1.25cm}{\includegraphics[width=4cm]{pentagon2.pdf}}\ \ \ =\sum_{i<j<k<l<i}Q_{ijkl}
\end{eqnarray}
where $i,j,k,l=1,\cdots,n$ and the notation $i<j<k<l<i$ in the sum is the cyclic sum.  \par
In this representation there are no spurious poles and this local representation of the amplitudes is corresponding to the BCFW \cite{localintegral}. Because of this corresponding, it is enough to compare with the local representation.  
 Next we construct the triangulation of the 2-loop MHV amplituhedron from the geometry using the sign flip definition and compare with the full form and the local representation from the double pentagon diagrams.

\section{Triangulation of 2-loop MHV Amplituhedron from Sign Flips}
In this section, we see that the sign flip pattern gives a natural triangulation of the amplituhedron. First we consider the tree $m=1, k=1$ case. From the definition of sign flips, $\{\langle Y1\rangle,\cdots,\langle Yn\rangle\}$ has $1$ sign flip. We denote the place where the sign flip takes place $j$; $\langle Yj\rangle<0$ and $\langle Yj+1\rangle>0$. Now we can expand $Y$ on some basis $\mathcal{Z}_A,\mathcal{Z}_B$ as $Y=\mathcal{Z}_A+x\mathcal{Z}_B$. In order to describe the $m = 1$ cell where the sign flip occurs at $j$, it is convenient to choose $\mathcal{Z}_A=\mathcal{Z}_{j}$, $\mathcal{Z}_B=\mathcal{Z}_{j+1}$. From the sign flip conditions, we must have $x_j>0$ and conversely, every $Y$ of this form with $x>0$ will belong to this cell. Then the canonical form for this sign flip pattern is 
\begin{equation}
\Omega_j=\frac{dx_j}{x_j}
\end{equation}
and the full form of $m=1$ $k=1$ amplituhedron is
\begin{equation}
\Omega=\sum_{1\leq j\leq n-1}\frac{dx_j}{x_j}.
\end{equation}
This is the triangulation of the $m=1, k=1$ tree amplituhedron from sign flips. We can similarly triangulate for general $k$. The region in the $m=1, k$ amplituhedron where $\{\langle Y1\rangle\cdots \langle Yn\rangle\}$  flips in slots $(j_1,\cdots j_k)$ is covered by
\begin{equation}
Y=(\mathcal{Z}_{j_1}+x_1\mathcal{Z}_{j_1+1})(\mathcal{Z}_{j_2}+x_2\mathcal{Z}_{j_2+1})\cdots (\mathcal{Z}_{j_k}+x_k\mathcal{Z}_{j_k+1})\ \ \text{with}\ x_k>0.
\end{equation}
The $k$-form related to each cell is
\begin{equation}
\Omega^{\{j_1,\cdots j_k\}}=\prod_{\alpha=1}^k d\log x_\alpha =\prod_{\alpha=1}^k d\log \frac{\langle Y\mathcal{Z}_{i_\alpha+1}\rangle}{\langle Y\mathcal{Z}_{i_\alpha}\rangle}
\end{equation}
and the full form is 
\begin{equation}
\Omega=\sum_{2\leq j_1\leq j_2\leq\cdots\leq j_k\leq n-1}\Omega^{\{j_1,\cdots j_k\}}.
\end{equation}
The dimension of each cell is $k$ and this is the triangulation of $\mathcal{A}(m=1,k,n)$ into non-redundant cells.
Similarly the $m=2$ amplituhedron can be triangulated from the sign flips. For $m=2$ case, the sequence $\{\langle Y12\rangle,\langle Y13\rangle,\cdots,\langle Y1n\rangle\}$ has $k$ sign flip in the slots $(j_1,\cdots j_k)$. Then we can expand $Y$ as 
\begin{equation}
\label{eq:expand}
Y=(+\mathcal{Z}_{1}+x_1\mathcal{Z}_{j_1}+y_1\mathcal{Z}j_1+1)(-\mathcal{Z}_{1}+x_2\mathcal{Z}_{j_2}+y_2\mathcal{Z}j_2+1)\cdots ((-1)^k\mathcal{Z}_{1}+x_k\mathcal{Z}_{j_k}+y_k\mathcal{Z}j_k+1)
\end{equation}
with $x_k,y_k>0$. The $k$-form for each cell is
\begin{equation}
\Omega^{\{j_1,\cdots j_k\}}=\prod_{\alpha=1}^k d\log x_\alpha d\log y_\alpha =\prod_{\alpha=1}^k d\log \frac{\langle Y1i_\alpha\rangle}{\langle Yi_\alpha i_\alpha+1\rangle}d\log \frac{\langle Y1i_\alpha+1\rangle}{\langle Yi_\alpha i_\alpha+1\rangle}
\end{equation}
and the full form is 
\begin{equation}
\Omega=\sum_{2\leq j_1\leq j_2\leq\cdots\leq j_k\leq n-1}\Omega^{\{j_1,\cdots j_k\}}.
\end{equation}
The dimension of each cell is $2k$ and this is the triangulation of $\mathcal{A}(m=2,k,n)$.  The important case is $k=2$, it is isomorphic to the 1-loop MHV amplituhedron $\mathcal{A}(k=0,n,m=4,l=1)$. The full canonical form of this amplituhedron $\mathcal{A}(2, n, 2)$ from sign flips is 
\begin{eqnarray}
\Omega&=&\sum_{2\leq j_1\leq j_2\leq\cdots\leq j_k\leq n-1}\Omega^{\{j_1,\cdots j_k\}}\nonumber\\
&=&\langle Y d^2Y_1 \rangle\langle Y d^2Y_2 \rangle \times\nonumber\\
&&\sum_{2\leq j_1\leq j_2\leq\cdots\leq j_k\leq n-1} \frac{\langle Y_1Y_2 (1ii+1)\cap(1jj+1)\rangle^2}{\langle Y1i\rangle\langle Y1i+1 \rangle \langle Y ii+1\rangle\langle Y1j\rangle\langle Y1j+1 \rangle \langle Y jj+1\rangle}.
\end{eqnarray}
Take $(Y_1,Y_2)\leftrightarrow(A,B)$, this form corresponds to the Kermit representation of 1-loop MHV amplituhedron, we have already known that this is the triangulation of this amplituhedron. \par

For $m=1$ and $m=2$ case, we have seen that the sign flip pattern gives us a triangulation of the amplituhedron and it comes from the geometry directly. However, for the case of $m=4$, there isn't a simple relation between the triangulation and the sign flip pattern. The case of the 1-loop MHV amplituhedron is an exception due to the isomorphism. However, we will see that we can triangulate the 2-loop MHV amplituhedron $\mathcal{A}(k=0,n,l=2)$ using the sign flip definition. 
The sign flip definition of the 2-loop MHV amplituhedron is 
\begin{eqnarray}
&&\langle AB i i+1 \rangle >0,\ \ \langle CD i i+1 \rangle >0\nonumber\\
&&\{\langle AB12 \rangle, \langle AB13 \rangle,\cdots,\langle AB 1n \rangle \} \ \ \text{has 2 sign flip}\nonumber\\
&&\{\langle CD12 \rangle, \langle CD13 \rangle,\cdots,\langle CD 1n \rangle \} \ \ \text{has 2 sign flip}\nonumber\\
&&\langle ABCD \rangle >0,
\end{eqnarray}
$(A,B)$ and $(C,D)$ are the loop momentum for each amplituhedron. From this, we can see that the 2-loop MHV amplituhedron is constructed by two 1-loop MHV amplituhedron $(AB)$, $(CD)$ and a further constraint $\langle ABCD \rangle>0$. The important fact is that even if we consider the general $n$-point, there is only one constraint $\langle ABCD \rangle>0$. Because of this, to obtain the canonical form we need to solve only one constraint and it is very easy rather than the $Y=C\cdot Z$ description. In the next section, we construct the triangulation of the 2-loop MHV amplituhedron from the sign flip definition and compare with the BCFW and the double pentagon representation. 

\subsection{Four and five point amplitudes}
First we consider the simplest case, 2-loop 4-point MHV amplituhedron. From the sign flip definition (\ref{eq:defloop}), it is constructed from the two 1-loop 4-point MHV. The sign flip definition of the 1-loop amplituhedron is
\begin{eqnarray}
&&\langle AB i i+1 \rangle >0,\nonumber\\
&&\{\langle AB12 \rangle, \langle AB13 \rangle,\langle AB 14 \rangle \} \ \ \text{has 1 sign flip}.
\end{eqnarray}
From this, there is only 1 sign flip pattern
\begin{equation}
\{ \langle AB12 \rangle ,\langle AB13 \rangle ,\langle AB14 \rangle \}=\{+,-,+\}.
\end{equation}
Then we can expand the loop momentum as  
\begin{equation}
 Z_A=Z_1+x_1Z_2+w_1Z_3,\ \ \  Z_B=-Z_1+y_1Z_3+z_1Z_4.
\end{equation}
From the sign flip condition, the region of these variables are $x_1,w_1,y_1,z_1>0$. In the view of the $Y_\alpha =C_{\alpha a}\mathcal{Z}_a$ description, the $C$-matrix of this sign flip pattern is 
\begin{equation}
C = \left(
    \begin{array}{cccc}
      1 & x_1 & w_1 & 0 \\
      -1 &0 & y_1 & z_1  
    \end{array}
  \right).
\end{equation}
Boundary of this pattern is $x_1\rightarrow0, w_1\rightarrow0,y_1\rightarrow0,z_1\rightarrow0$, then the canonical form is 
\begin{eqnarray}
\Omega^{l=1}_4=\frac{dx_1}{x_1}\frac{dw_1}{w_1}\frac{dy_1}{y_1}\frac{dz_1}{z_1}=\frac{\langle ABd^2A \rangle \langle AB d^2 B\rangle \langle 1234 \rangle^2}{\langle AB12 \rangle \langle AB23 \rangle \langle AB34 \rangle \langle AB14 \rangle}.
\end{eqnarray}
This form corresponds to the form of the 4-point 1-loop MHV amplituhedron obtained from the $Y=C\cdot Z$ description \cite{amplituhedron,intoamplituhedron}. 
Next we consider the 2-loop 4-point MHV amplituhedron. This is constructed from the two 1-loop amplituhedron and a constraint $\langle ABCD \rangle >0$. We can parametrize these two 1-loop amplituhedron as
 \begin{eqnarray}
 \label{eq:2323parameter}
 &Z_A=Z_1+x_1Z_2+w_1Z_3,\ \ \  Z_B=-Z_1+y_1Z_3+z_1Z_4 \nonumber\\
  &Z_C=Z_1+x_2Z_2+w_2Z_3,\ \ \  Z_D=-Z_1+y_2Z_3+z_2Z_4\nonumber\\
  &\text{with} \ \ \ x_1,w_1,y_1,z_1,x_2,w_2,y_2,z_2>0.
  \end{eqnarray}
  In view of the $\mathcal{Y}=\mathcal{C}\cdot Z$ description, the $C$-matrix is
\begin{equation}
\label{eq:4ptcmatrix}
C = \left(
    \begin{array}{cccc}
      1 & x_1 & w_1 & 0 \\
      -1 &0 & y_1 & z_1  \\
      1 & x_2 & w_2 & 0 \\
      -1 &0 & y_2 & z_2
    \end{array}
  \right).
\end{equation}
Under this parametrization, the constraint become
\begin{equation}
\langle ABCD\rangle=\langle 1234 \rangle \{(x_1-x_2)(y_1z_2-y_2z_1)+(z_1-z_2)(w_1x_2-w_2x_1)\}>0.
\end{equation}
From this condition, these parameters are bounded further. Without loss of generality, we can take $y_1z_2-y_2z_1>0$. Then from $\langle ABCD \rangle >0$, 
\begin{equation}
x_1>x_2-\frac{(z_1-z_2)(w_1x_2-w_2x_1)}{y_1z_2-y_2z_1}=x_2-a.
\end{equation}
Therefore there are 4 cases depending on the signs of $( z_1-z_2),\ (w_1x_2-w_2x_1)$.  For example, the case of $(z_1-z_2)>0, \ (w_1x_2-w_2x_1)>0$, the regions of these variables are 
\begin{eqnarray}
x_2+a>x_1>0,\ \ w_1>\frac{x_1}{x_2}w_2,\ \ y_1>\frac{z_1}{z_2}y_2,\ \ z_1>z_2\nonumber\\
x_2>0,\ \ w_2>0,\ \ y_2>0,\ \ z_2>0.
\end{eqnarray}
Compare with (\ref{eq:2323parameter}), the regions of these parameters are further bounded
because of this constraint. Then there are 9 boundaries
\begin{eqnarray}
&(x_1\rightarrow x_2+a,\ \ x_1\rightarrow0),\ \ w_1\rightarrow\frac{x_1}{x_2}w_2,\ \ y_1\rightarrow\frac{z_1}{z_2}y_2,\ \ z_1\rightarrow z_2\nonumber\\
&x_2\rightarrow0,\ \ w_2\rightarrow0,\ \ y_2\rightarrow0,\ \ z_2\rightarrow0.
\end{eqnarray}
We can obtain the canonical form for this case. For example, the region of $x_1$ is $0<x_1<x_2+a$, then the form for $x_1$ is
\begin{equation}
\frac{1}{x_1}-\frac{1}{x_1-x_2-a}.
\end{equation}
Then the canonical form for this case is
\begin{equation}
\Omega=\frac{1}{x_2}\left(\frac{1}{x_1}-\frac{1}{x_1-x_2-a}\right)\frac{1}{w_1-\frac{x_1}{x_2}w_2}\frac{1}{w_2}\frac{1}{y_1-\frac{z_1}{z_2}y_2}\frac{1}{y_2}\frac{1}{z_1-z_2}\frac{1}{z_2}.
\end{equation}
There are 4 patterns depending on the signs of $( z_1-z_2),\ (w_1x_2-w_2x_1)$.  The forms related to these 4 patterns can be constructed similarly 
\begin{eqnarray}
\Omega_1&=&\frac{1}{x_1-x_2+a}\frac{1}{x_2}\left(\frac{1}{w_1}-\frac{1}{w_1-\frac{x_1}{x_2}w_2}\right)\frac{1}{w_2}\frac{1}{y_1-\frac{z_1}{z_2}y_2}\frac{1}{y_2}\frac{1}{z_1-z_2}\frac{1}{z_2}\nonumber\\
\Omega_2&=&\frac{1}{x_1-x_2+a}\frac{1}{x_2}\frac{1}{w_1-\frac{x_1}{x_2}w_2}\frac{1}{w_2}\frac{1}{y_1-\frac{z_1}{z_2}y_2}\frac{1}{y_2}\left(\frac{1}{z_1}-\frac{1}{z_1-z_2}\right)\frac{1}{z_2}\nonumber\\
\Omega_3&=&\frac{1}{x_1}\left(\frac{1}{x_2}-\frac{1}{x_2-x_1-a}\right)\frac{1}{w_1-\frac{x_1}{x_2}w_2}\frac{1}{w_2}\frac{1}{y_1-\frac{z_1}{z_2}y_2}\frac{1}{y_2}\frac{1}{z_1-z_2}\frac{1}{z_2}\\
\Omega_4&=&\frac{1}{x_1}\left(\frac{1}{x_2}-\frac{1}{x_2-x_1-a}\right)\left(\frac{1}{w_1}-\frac{1}{w_1-\frac{x_1}{x_2}w_2}\right)\frac{1}{w_2}\frac{1}{y_1-\frac{z_1}{z_2}y_2}\frac{1}{y_2}\left(\frac{1}{z_1}-\frac{1}{z_1-z_2}\right)\frac{1}{z_2}\nonumber.
\end{eqnarray}
The remaining four cases $y_1z_2-y_2z_1<0$ are obtained that swap $1 \leftrightarrow 2$.
The sum of these 8 form is
\begin{equation}
\Omega^{l=2}_{4\text{pt}}=\frac{dx_1dx_2dw_1dw_2dy_1dy_2dz_1dz_2}{x_1x_2w_1w_2y_1y_2z_1z_2}\frac{(x_1y_1z_2+x_2y_2z_1+x_2w_1z_1+x_1w_2z_2)}{\{(x_1-x_2)(y_1z_2-y_2z_1)+(z_1-z_2)(w_1x_2-w_2x_1)\}}.
\end{equation}
To translate it into the momentum twistor, we need to solve (\ref{eq:2323parameter}) for $x_1,x_2,\cdots,z_2$ 
\begin{eqnarray}
x_1=-\frac{\langle AB13\rangle}{\langle AB23\rangle},\ \ w_1=\frac{\langle AB12\rangle}{\langle AB23\rangle},
y_1=\frac{\langle AB14\rangle}{\langle AB34\rangle},\ \ z_1=-\frac{\langle AB13\rangle}{\langle AB34\rangle}\nonumber\\
x_2=-\frac{\langle CD13\rangle}{\langle CD23\rangle},\ \ w_2=\frac{\langle CD12\rangle}{\langle CD23\rangle},
y_2=\frac{\langle CD14\rangle}{\langle CD34\rangle},\ \ z_2=-\frac{\langle CD13\rangle}{\langle CD34\rangle}.
\end{eqnarray}
Then the full form in the momentum twistor space is
\begin{eqnarray}
\label{eq:4ptform}
\!\!\!\!\!\!\!\!\!\!\!\!\!\!\!\!\!\!\Omega^{l=2}_{4\text{pt}}&=&\frac{\langle 1234 \rangle^3 \langle ABd^2A \rangle \langle ABd^2B\rangle \langle CDd^2C \rangle \langle CDd^2D\rangle  }{\langle AB12 \rangle \langle AB14 \rangle\langle AB23 \rangle \langle AB34 \rangle\langle ABCD \rangle\langle CD12 \rangle \langle CD14 \rangle\langle CD23 \rangle \langle CD34 \rangle}\nonumber\\
&\times&\biggr\{ \langle AB34 \rangle \langle CD12 \rangle+\langle AB23 \rangle \langle CD14 \rangle+\langle AB14 \rangle \langle CD23 \rangle+\langle AB12 \rangle \langle CD34 \rangle\biggl\}.
\end{eqnarray}
This result is corresponding to the local representation of the 4-point amplitude (\ref{eq:local4pt}). The dimension of this amplituhedron is $8$, therefore in this 4-point case, it is just a non-redundant cell. Of cause it can be obtained from the $Y=C\cdot Z$ description directly \cite{intoamplituhedron} and our result is corresponding to this $Y=C\cdot Z$ result. Next we see that the higher point 2-loop MHV amplituhedron can be triangulated into the non-redundant dimension $8$ cells.\par
\ \par
Next we consider the 5-point amplitude. The 2-loop 5-point MHV amplituhedron is constructed from the two 1-loop 5-point MHV amplituhedron and a further constraint. In the 1-loop $n=5,k=2$ amplitude, there are 3 patterns of sign flips as
\begin{equation}
\{\langle AB 12 \rangle,\langle AB 13 \rangle,\langle AB 14 \rangle,\langle AB 15 \rangle\}=\{+,-,+,+\}\ \ \text{or}\ \ \{+,-,-,+\}\ \ \text{or}\ \ \{+,+,-,+\}.
\end{equation}
Then we can parametrize for each pattern as
\begin{eqnarray}
\label{eq:5ptpattern}
&&\begin{cases}
 Z_A=Z_1+x_1Z_2+w_1Z_3\\ Z_B=-Z_1+y_1Z_3+z_1Z_4\nonumber
\end{cases}
(2,3)\text{\ pattern}\ ,
\begin{cases}
Z_A=Z_1+x_1Z_2+w_1Z_3\\ Z_B=-Z_1+y_1Z_4+z_1Z_5\nonumber
\end{cases}
(2,4)\text{\ pattern},\\
&&\begin{cases}
Z_A=Z_1+x_1Z_3+w_1Z_4\\ Z_B=-Z_1+y_1Z_4+z_1Z_5
\end{cases}
(3,4)\text{\ pattern}.
\end{eqnarray}
Then depending on which pattern \eqref{eq:5ptpattern} we choose, there are $3\times3=9$ patterns in the 2-loop amplituhedron. We can expect that the full form of the 2-loop 5-point MHV amplituhedron is obtained by the sum of these forms related to each 9 pattern. Each form can be obtained similarly as the 4-pt case, and the explicit calculation is given in the appendix and here we will write only the results.
The case of $(2,3)\times(2,3)$ is same as the 4-pt case.
The case of $(3,4)\times(3,4)$, 
\begin{eqnarray}
\Omega_{3434}&=&\frac{dx_1dx_2dw_1dw_2dy_1dy_2dz_1dz_2}{x_1x_2w_1w_2y_1y_2z_1z_2}\frac{\langle 1345 \rangle}{\langle ABCD \rangle}(x_1y_1z_2+x_2y_2z_1+x_2w_1z_1+x_1w_2z_2)\nonumber\\
&=&\frac{\langle 1345 \rangle^3 \langle ABd^2A \rangle \langle ABd^2B\rangle \langle CDd^2C \rangle \langle CDd^2D\rangle}{\langle AB13 \rangle \langle AB15 \rangle\langle AB34 \rangle \langle AB45\rangle\langle ABCD \rangle\langle CD13 \rangle \langle CD15 \rangle\langle CD34 \rangle \langle CD45 \rangle}\nonumber\\
&\times&\biggr\{ \langle AB45 \rangle \langle CD13 \rangle+\langle AB34 \rangle \langle CD15 \rangle+\langle AB15 \rangle \langle CD34 \rangle+\langle AB13 \rangle \langle CD45 \rangle\biggl\}.\ \ \ \ \ \ \ \ \ 
\end{eqnarray}
The case of $(2,4)\times(3,4)$, 
\begin{eqnarray}
\Omega_{2434}&=&\frac{\langle 123A_4\rangle\langle 134C_4 \rangle \langle ABd^2A \rangle \langle ABd^2B\rangle \langle CDd^2C \rangle \langle CDd^2D\rangle}{\left\{\LARGE{\substack{\langle AB12 \rangle\langle AB13\rangle\langle AB14 \rangle\langle AB15 \rangle  \langle AB23\rangle\langle AB45\rangle\\\times\langle ABCD \rangle\langle CD13 \rangle\langle CD14\rangle^2\langle CD15\rangle\langle CD34\rangle\langle CD45\rangle}}\right\}}\nonumber
\\
&\times&\biggl\{\langle 123 A_4\rangle(\langle AB45 \rangle\langle CD13 \rangle\langle CD14\rangle+\langle AB15 \rangle\langle CD34 \rangle\langle CD14\rangle)\nonumber\\
&&\ \ -\langle 345 A_2\rangle\langle AB14 \rangle\langle CD14\rangle\langle CD15 \rangle+\langle 123C_4\rangle\langle CD14\rangle\langle AB45 \rangle\langle AB13 \rangle\biggr\}.\ \ 
\end{eqnarray}
\\
The case of $(2,3)\times(3,4)$, 
\begin{eqnarray}
\Omega_{2334}&=&\frac{\langle 123A_3 \rangle \langle 134C_4\rangle\langle ABd^2A \rangle \langle ABd^2B\rangle \langle CDd^2C \rangle \langle CDd^2D\rangle}{\left\{\LARGE{\substack{\langle AB12 \rangle\langle AB13 \rangle^2\langle AB14\rangle\langle AB23\rangle\langle AB34\rangle\\\times\langle ABCD \rangle\langle CD13 \rangle\langle CD14\rangle^2\langle CD15\rangle\langle CD34\rangle\langle CD45\rangle}}\right\}}\nonumber
\\
&\times&\biggl\{\langle AB13 \rangle\langle 123 C_4\rangle\langle CD4 A_3\rangle-\langle AB13 \rangle\langle AB14\rangle\langle CD13\rangle\langle 234 C_4\rangle\nonumber\\
&&\ \ +\langle CD14 \rangle\langle 145 A_2\rangle\langle CD3 A_3\rangle-\langle AB14 \rangle\langle AB23\rangle\langle CD13\rangle\langle CD14\rangle\langle1345 \rangle\biggr\}\nonumber.\\
\end{eqnarray}
The case of $(2,4)\times(2,4)$, 
\begin{eqnarray}
\Omega_{2424}&=&\frac{\langle 123A_4 \rangle \langle 123C_4\rangle\langle ABd^2A \rangle \langle ABd^2B\rangle \langle CDd^2C \rangle \langle CDd^2D\rangle}{\left\{\LARGE{\substack{\langle AB12 \rangle\langle AB13\rangle\langle AB14\rangle\langle AB15\rangle\langle AB23\rangle\langle AB45\rangle\langle ABCD \rangle\\\times\langle CD12 \rangle\langle CD13\rangle\langle CD14\rangle\langle CD15\rangle\langle CD23\rangle\langle CD45\rangle}}\right\}}\nonumber
\\
&\times&\biggl\{\langle 123 A_4\rangle(\langle AB12 \rangle\langle CD13 \rangle\langle CD45\rangle+\langle AB15 \rangle\langle CD14 \rangle\langle CD23 \rangle)\nonumber\\
&&+\langle 123 C_4\rangle(\langle AB13 \rangle\langle AB45 \rangle\langle CD12 \rangle+\langle AB14 \rangle\langle AB23 \rangle\langle CD15 \rangle)\nonumber\\
&&+\langle 2345 \rangle(\langle AB12 \rangle\langle AB15 \rangle\langle CD13 \rangle\langle CD14 \rangle+\langle AB13 \rangle\langle AB14 \rangle\langle CD12 \rangle\langle CD15 \rangle)\biggr\}\nonumber.\\
\end{eqnarray} 
We use the symbols that
 \begin{eqnarray}
 \label{eq:symbol}
A_i&\equiv&(AB)\cap(1ii+1),\ \ C_k\equiv (CD)\cap(1 kk+1).
\end{eqnarray}
The remaining patterns are $(3,4)\times(2,3), (2,4)\times(2,3),(3,4)\times(2,4)$. These forms can be obtained from $\Omega_{2334},\Omega_{2324},\Omega_{2434}$ that swap $AB \leftrightarrow CD$. We obtain all 9 forms and we can calculate the sum of these forms
\begin{equation}
\Omega_{5\text{-pt}}^{l=2, \text{MHV}}=\Omega_{2323}+\Omega_{2424}+\Omega_{3434}+\Omega_{2324}+\Omega_{2334}+\Omega_{2434}+\Omega_{2423}+\Omega_{3423}+\Omega_{3424}.
\end{equation}
 Each form $\Omega_{ijkl}$ has spurious poles $\langle AB13\rangle, \langle AB14 \rangle, \langle CD13 \rangle, \langle CD14 \rangle$, we can see that all of these are canceled and remain only the physical poles in the full form. This result is corresponding to the local representation (\ref{eq:local5pt}).  From this result, we can see that the 2-loop 5-point MHV amplituhedron is triangulated into the 9 cells related to each sign flip pattern, and these cells are 8-dimensional cells $G_+(4,4)$. \par
 In the case of the BCFW, each cell of the 2-loop 5-point MHV amplitude has also the spurious poles not only like $\langle AB13\rangle, \langle AB14 \rangle, \langle CD13 \rangle, \langle CD14 \rangle$, but also more complicate poles from taking the forward limit. Therefore this triangulation has a different structure compared with the BCFW triangulation.

\subsection{n-point amplitude}
Next we consider the general n-pt case. First we consider the 1-loop n-point MHV amplituhedron. There are $\frac{1}{2}(n-3)(n-2)$ sign flip patterns from the way to chose $i,j$ that
\begin{equation}
i,j=2,3,\cdots,n-1,\ \ \ i<j.
\end{equation}
When sign flip occurs at $i,j$ slots, we can parametrize the loop momentum as
\begin{equation}
Z_A=Z_1+xZ_i+wZ_{i+1},Z_B=-Z_1+yZ_j+zZ_{j+1},
\end{equation}
and the canonical form of this pattern is
\begin{equation}
\Omega_{ij}=\frac{dx}{x}\frac{dw}{w}\frac{dy}{y}\frac{dz}{z}.
\end{equation}
Then the full form is
\begin{equation}
\Omega=\sum_{\substack{i,j=2,3,\cdots,n-1\\i<j}}\Omega_{ij}.
\end{equation}
Next we consider the 2-loop $n$-point MHV amplituhedron. There are $[\frac{1}{2}(n-3)(n-2)]^2$ sign flip patterns in the 2-loop n-point MHV amplituhedron depending on the way to chose $i,j,k,l$ that
\begin{equation}
i,j,k,l=2,3,\cdots,n-1,\ \ \ i<j, k<l.
\end{equation}
We can expand as
\begin{eqnarray}
\label{eq:expand}
\begin{cases}Z_A=Z_1+x_1Z_i+w_1Z_{i+1},\\ Z_B=-Z_1+y_1Z_j+z_1Z_{j+1}\nonumber\end{cases}
\begin{cases}Z_C=Z_1+x_2Z_k+w_2Z_{k+1},\\Z_D=-Z_1+y_2Z_l+z_2Z_{l+1}\end{cases}.
\end{eqnarray}
From the constraint $\langle ABCD \rangle>0$, these parameters are bounded. The region of these parameters are depending on the other parameters and $\langle ijkl \rangle$,  however, the sign of this determinant changes depending on the relation between $(i,j)$ and $(k,l)$.  
More precisely, the sign is depending on the order of $i,j,k,l$, if $i<j<k<l$, then $\langle ijkl\rangle >0$. Therefore we need to determine the order of $i,j,k,l$ to calculate each form. This order of $i,j,k,l$ can be divided into 13 groups as
\begin{eqnarray}
\label{eq:pattern}
i<k<l<j\cdots(1),&&i<k<j<l\cdots(2), \ \ i<j<k<l\cdots(3), \ \ i=k<l<j\cdots(4),\nonumber\\
i=k<j=l\cdots(5),&&i=k<j<l\cdots(6), \ \ i<k<j=l\cdots(7),\ \ i<j=k<l\cdots(8),\nonumber\\
k<i=l<j\cdots(9),&&k<i<l<j\cdots(10), \ \ k<i<j=l\cdots(11), \ \ k<i<j<l\cdots(12),\nonumber\\
k<l<i<j\cdots(13).&&
\end{eqnarray}
We can compute the forms for each case in the same way as the 5-point case. The case of  $(1)$, $C$-matrix is
\begin{equation}
C = \left(
    \begin{array}{cccccccccc}
      1  &\cdots & i&\cdots&\cdots&\cdots&\cdots&\cdots&\cdots&\cdots\\
      -1 &\cdots&\cdots&\cdots&\cdots&\cdots&\cdots&\cdots& j&\cdots\\
      1 &\cdots&\cdots&\cdots&k&\cdots &\cdots&\cdots&\cdots&\cdots  \\
      -1 &\cdots&\cdots&\cdots&\cdots&\cdots&l &\cdots &\cdots&\cdots
    \end{array}
  \right)
\end{equation}
where
\begin{equation}
(i,i+1\rightarrow x_1,w_1)\ \ (j,j+1\rightarrow y_1,z_1)\ \ (k,k+1\rightarrow x_2,w_2)\ \ (l,l+1\rightarrow y_2,z_2)\ \ ,\cdots = 0.
\end{equation}
The canonical form of this case in the momentum twistor space is
 \begin{equation}
\Omega^1_{ijkl}=\frac{\omega^{1'}_{ijkl}\langle 1ii+1 A_j \rangle\langle1kk+1 C_l \rangle\langle AB d^2A \rangle\langle AB d^2B \rangle\langle CD d^2C \rangle\langle CD d^2D \rangle}{\langle AB 1i \rangle\langle AB 1i+1 \rangle\langle AB 1j \rangle\langle AB 1j+1 \rangle\langle ABCD \rangle\langle CD 1k \rangle\langle CD 1k+1 \rangle\langle CD1l \rangle\langle CD 1l+1 \rangle}
\end{equation}
where
  \begin{equation}
\omega_{ijkl}^{1'}=\frac{\langle ABii+1 \rangle\langle A_j C_k C_l 1\rangle+\langle A_iA_jC_k C_l  \rangle}{\langle ABii+1\rangle\langle ABjj+1\rangle\langle CDkk+1\rangle\langle CDll+1\rangle}.
\end{equation}
 Again we use the symbols (\ref{eq:symbol}).
The canonical forms for another case can be obtained similarly. We give all the canonical forms and the explicit calculation of the case of (1) in the appendix.\par
Then the full form of the 2-loop n-pt MHV amplituhedron is 
\begin{equation}
\Omega^{n\text{-pt 2-loop}}_{\text{MHV}}=\sum_{\substack{i,j,k,l=2,3,\cdots,n-1\\i<k<l<j}}\Omega_{ijkl}^1+\sum_{i<k<j<l}\Omega_{ijkl}^2+\sum_{i<j<k<l}\Omega_{ijkl}^3+\cdots+\sum_{k<l<i<j}\Omega_{ijkl}^{13}.
\end{equation}
Similarly for the 5-pt case, these cells have spurious poles. However, all of these poles are canceled and remain only physical poles. We compared this result and local representation (\ref{eq:localnpt}) (or BCFW) numerically \cite{all2loop} and we checked that these results are corresponding up to at least 22-pt. From this results, we can see that the 2-loop n-pt MHV amplituhedron is triangulated into the $[\frac{1}{2}(n-3)(n-2)]^2$ 8-dimension cells and this triangulation is obtained directly from the geometry. 

\section{More 2-loop Objects}
\subsection{Log of the 2-loop MHV Amplitude}
In this section we consider the log of the 2-loop MHV amplitude. The expansion of the amplitude is 
\begin{equation}
\mathcal{A}=1+gA_1+g^2A_2+g^3A_3+\cdots.
\end{equation}
Then the expansion of the logarithm of the amplitude is
\begin{equation}
\mathcal{S}=\log{\mathcal{A}}=gS_1+g^2S_2+g^3S_3+\cdots
\end{equation}
where $S_L$ is a sum of $A_L$ and products of lower-loop amplitude,
\begin{equation}
\label{eq:logamplitude}
S_1=A_1,\ \ \ S_2=A_2-\frac{1}{2}A_1^2,\ \ \ S_3=A_3-A_2A_1+\frac{1}{3}A_1^3,\ \ \cdots,
\end{equation}
therefore the first non-trivial part is the 2-loop log amplitude. The 2-loop log amplitude can be expressed simply as a non-planar cyclic sum of the double pentagon diagram because of the simple relation between the square of the 1-loop pentagon diagram and the 2-loop double pentagon diagram \cite{localintegral}. The 1-loop pentagon diagram is
\begin{eqnarray}
  \raisebox{-1.25cm}{\includegraphics[width=3cm]{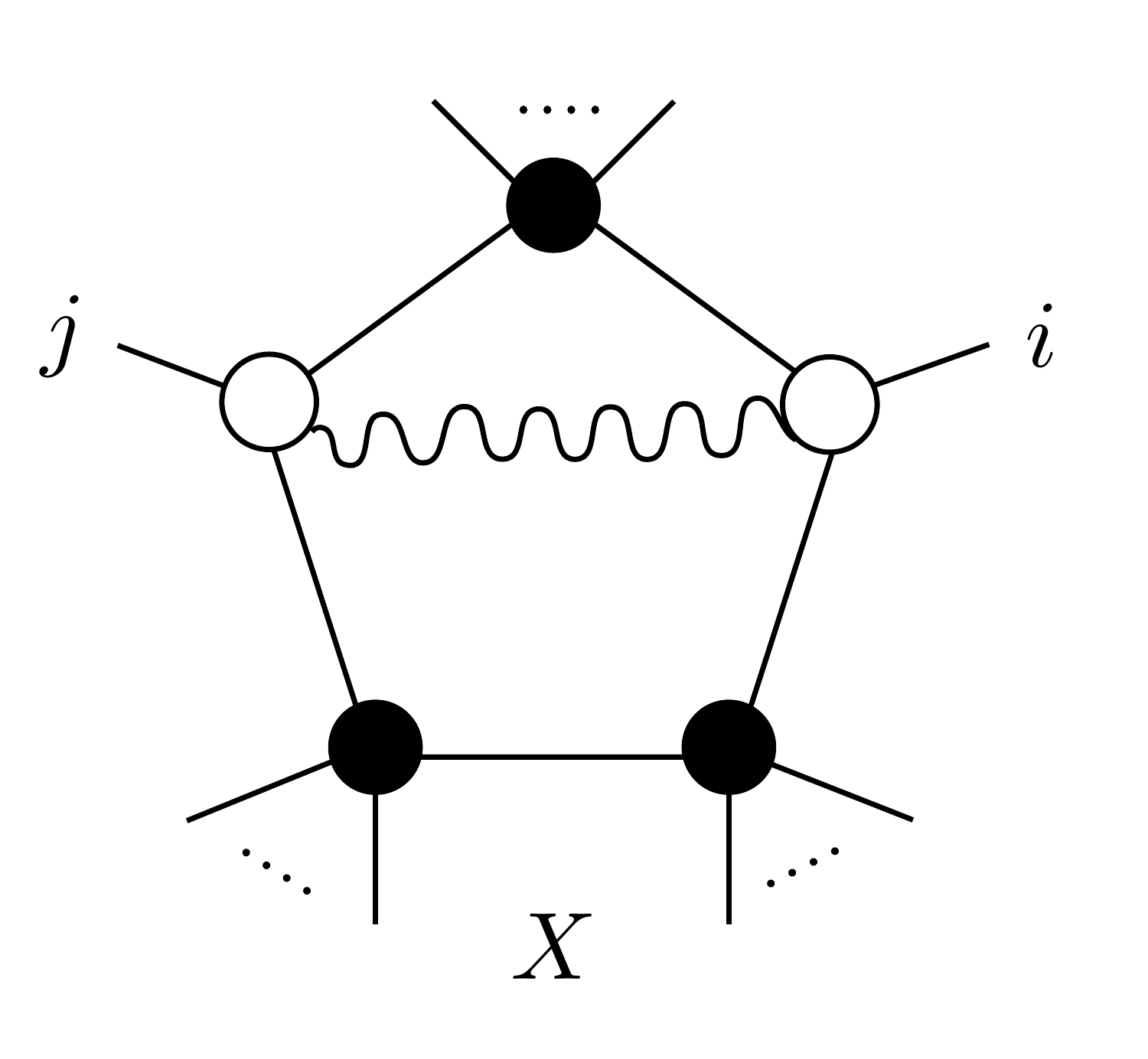}}=\frac{\langle AB(i-1ii+1)\cap(j-1jj+1)\rangle \langle Xij \rangle}{\langle ABX\rangle\langle ABi-1i\rangle\langle ABii+1\rangle\langle ABj-1j\rangle\langle ABjj+1\rangle}
\end{eqnarray}
and the 1-loop MHV amplitude is 
\begin{equation}
\mathcal{A}_{\text{MHV}}^{1-\text{loop}}=\sum_{i<j}\left\{\raisebox{-1.25cm}{\includegraphics[width=3cm]{pentagon.pdf}}\right\}.
\end{equation}
The relation between this pentagon diagram and the double pentagon diagram is
\begin{equation}
\sum_{i<j}\raisebox{-1.25cm}{\includegraphics[width=3cm]{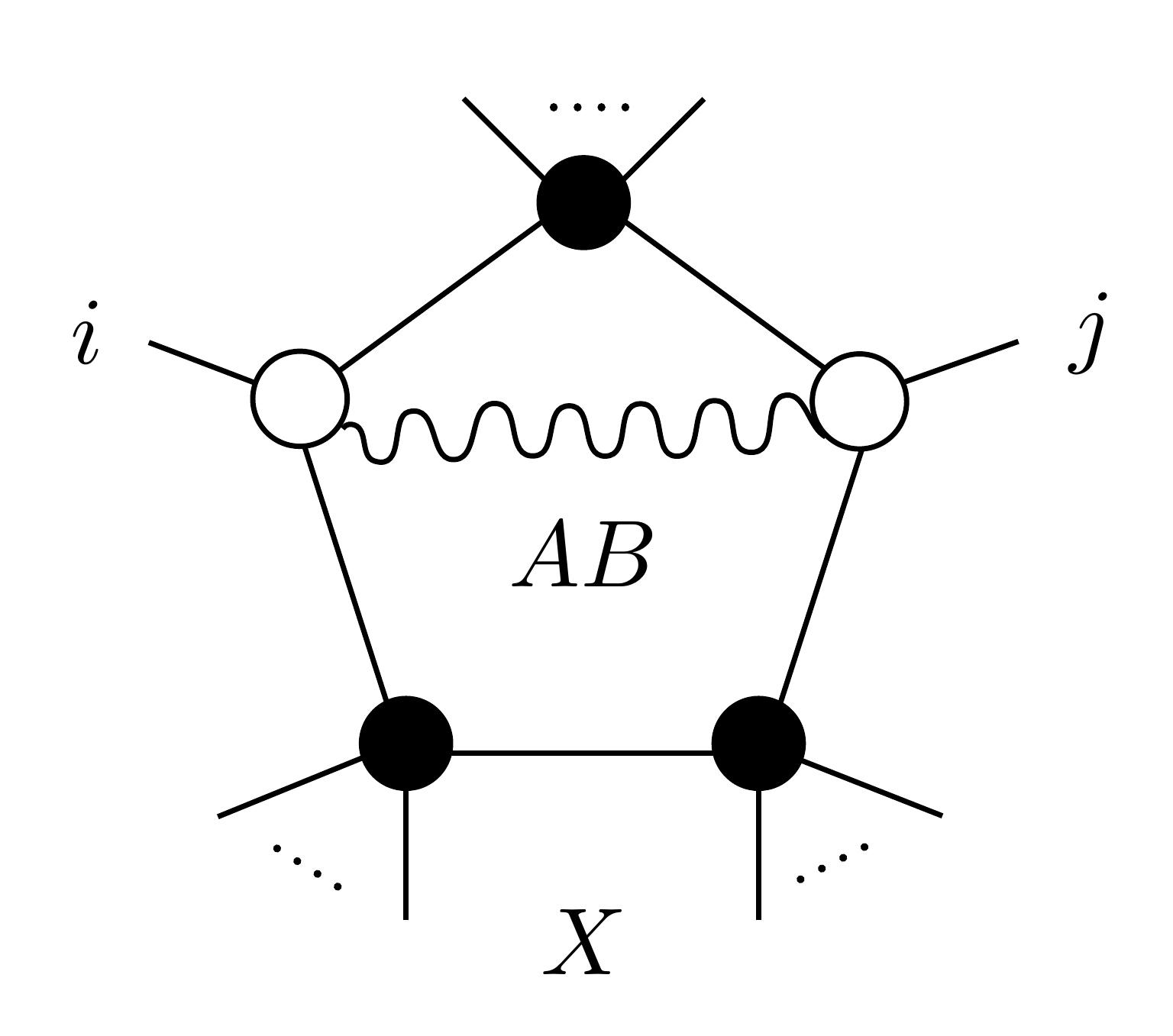}}\times\sum_{k<l}\raisebox{-1.25cm}{\includegraphics[width=3cm]{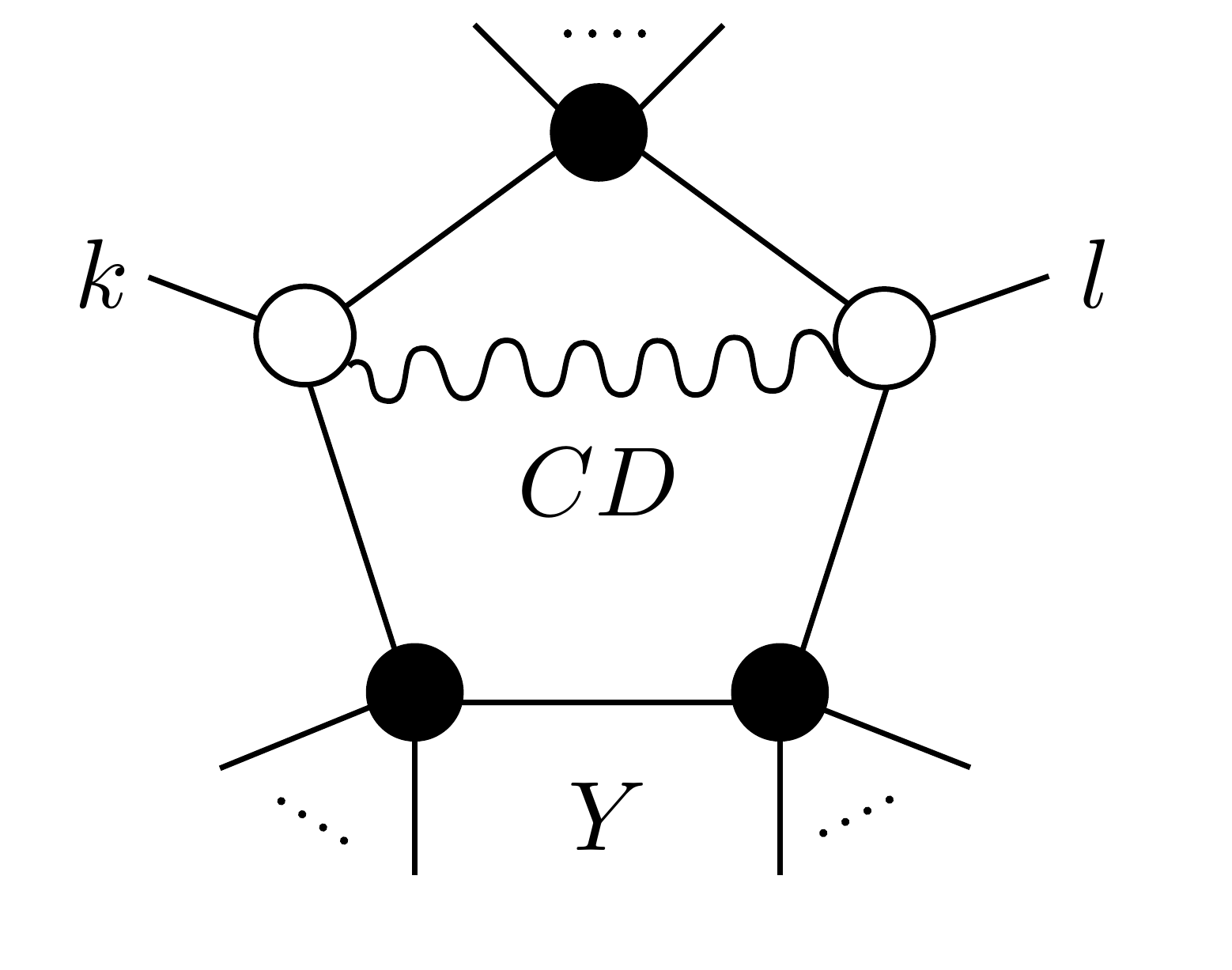}}=\sum_{i<j,k<l} \raisebox{-1.25cm}{\includegraphics[width=4cm]{pentagon2.pdf}}.
\end{equation}
The left side is just $(\mathcal{A}_{\text{MHV}}^{1-\text{loop}})^2$ and the right side contains not only the planar diagrams $i<j<k<l$ but also the non-planar diagrams; for example, $i<k<j<l$. From (\ref{eq:logamplitude}), the log of the 2-loop amplitude is 
\begin{equation}
\label{eq:logamplitude2}
-[\log{\mathcal{A}}]_{\text{MHV}}^{2-\text{loop}}=\frac{1}{2}(\mathcal{A}_{\text{MHV}}^{1-\text{loop}})^2-\mathcal{A}_{\text{MHV}}^{2-\text{loop}}.
\end{equation}
This means that the sum of all non-planar double pentagon diagrams times minus sign gives us the log of the 2-loop amplitude \cite{localintegral}
\begin{equation}
\label{eq:nonplanarsum}
[\log{\mathcal{A}}]^{2-\text{loop}}_{\text{MHV}}=-\sum_{i<k<j<l<i} \raisebox{-1.25cm}{\includegraphics[width=4cm]{pentagon2.pdf}}=-\sum_{i<k<j<l<i} Q_{ijkl}.
\end{equation}
For example, the log of the 4-pt amplitude is
\begin{eqnarray}
\label{eq:4ptlog}
&&[\log{\mathcal{A}}]^{2-\text{loop},\ 4\text{-pt}}_{\text{MHV}}=-Q_{1324}\nonumber\\
&=&\frac{\langle 1234 \rangle^3(\langle AB13 \rangle \langle CD24 \rangle+\langle AB24 \rangle \langle CD13 \rangle)}{\langle AB12 \rangle \langle AB23 \rangle\langle AB34 \rangle \langle AB14 \rangle\langle ABCD \rangle \langle CD12 \rangle\langle CD23 \rangle \langle CD34 \rangle\langle CD14 \rangle}.
\end{eqnarray}

Next we consider the log of the amplitude from the geometrical view. First we consider the region of the log of the amplitude. In the case of the 2-loop MHV, the definition of the amplituhedron is 
\begin{equation}
\mathcal{L}_{i}=D_{i}\cdot Z,\ \ \ i=1,2
\end{equation}
and
\begin{equation}
\langle D_{(1)}D_{(2)} \rangle>0.
\end{equation}
We consider another case: the square of the 1-loop MHV amplituhedron. This is defined similarly 
\begin{equation}
\label{eq:square}
\mathcal{L}_{i}=D_{i}\cdot Z,\ \ \ i=1,2.
\end{equation}
However, this has no positivity condition. From (\ref{eq:logamplitude2}), the region of the minus log of the amplitude is $\langle D_{(1)}D_{(2)} \rangle<0$. This pattern can be extended to all higher loop \cite{intoamplituhedron}. Then the question is that is it possible to obtain the log of the 2-loop MHV amplitude from the geometry
\begin{equation}
\label{eq:loggeometry}
\mathcal{L}_{i}=D_{i}\cdot Z,\ \ \ i=1,2,\ \ \ \text{and}\ \ \ \langle D_{(1)}D_{(2)} \rangle<0
\end{equation}
and the canonical form of this geometry? In this section we construct the canonical form of this space (\ref{eq:loggeometry}) and see that it is corresponding to the log of the 2-loop MHV amplitude. To obtain the canonical form, we use the sign flip definition of this geometry 
\begin{eqnarray}
\label{eq:logspace}
&&\langle AB i i+1 \rangle >0,\ \ \langle CD i i+1 \rangle >0\nonumber\\
&&\{\langle AB12 \rangle, \langle AB13 \rangle,\cdots,\langle AB 1n \rangle \} \ \ \text{has 2 sign flip}\nonumber\\
&&\{\langle CD12 \rangle, \langle CD13 \rangle,\cdots,\langle CD 1n \rangle \} \ \ \text{has 2 sign flip}\nonumber\\
&&\langle ABCD \rangle <0
\end{eqnarray}
and call this geometry "2-loop MHV log amplituhedron".
From this definition, we can see that the 2-loop n-point MHV log amplituhedron is constructed from the two 1-loop MHV amplituhedron and a negative constraint $\langle ABCD \rangle <0$. Then there are $[\frac{1}{2}(n-3)(n-2)]^2$ sign flip patterns in the 2-loop n-point MHV log amplituhedron depending on the way to chose $i,j,k,l$ that
\begin{equation}
i,j,k,l=2,3,\cdots,n-1,\ \ \ i<j, k<l.
\end{equation}
We can expand the loop momentum $(Z_A,Z_B), (Z_C,Z_D)$ as (\ref{eq:expand}) and the order of $i,j,k,l$ is divided into 13 groups as (\ref{eq:pattern}). Once we get the order of $i,j,k,l$, then we can calculate the canonical form similarly. For example, the canonical form for the case of (1) is
\begin{equation}
\Omega^{1}_{ijkl}[\log]=\frac{dx_1dx_2\cdots dz_1dz_2}{x_1x_2w_1w_2y_1y_2z_1z_2}\frac{-1}{(az_2-bw_1-cx_1-dw_2+ey_2)}\times\omega^1_{ijkl}[\log]
\end{equation}
where
\begin{eqnarray}
\omega^1_{ijkl}[\log]&=&x_2 y_1  \langle 1  i+1  k  j \rangle  + x_2 z_1  \langle 1  i+1  k  j+1 \rangle  + 
 y_1 y_2  \langle 1  i+1  l  j \rangle  + y_2 z_1  \langle 1  i+1  l  j+1 \rangle \nonumber \\&+& 
 y_1 z_2  \langle 1  i+1  l+1  j \rangle+ z_1 z_2  \langle 1  i+1  l+1  j+1 \rangle+ x_2 y_1 \langle 1  i  k  j\rangle  + x_2 z_1 \langle 1  i  k  j+1\rangle  \nonumber\\ &+& 
 w_2 y_1 \langle 1  i  k+1  j\rangle  + w_2 z_1 \langle 1  i  k+1  j+1\rangle  + 
 y_1 y_2 \langle 1  i  l  j\rangle  + y_2 z_1 \langle 1  i  l +1  j\rangle      \nonumber\\&+& 
 y_1 z_2 \langle 1  i  l+1  j\rangle  + z_1 z_2 \langle 1  i  l+1  j+1\rangle +w_1 y_1 \langle 1  i+1  k+1  j\rangle      \nonumber\\&+& w_1 z_1 \langle 1  i+1  k+1  j+1\rangle
\end{eqnarray}
and 
\begin{eqnarray}
\label{eq:abcddef}
a&= &x_1  x_2 \langle 1  i  k  l+1\rangle +  w_2  x_1 \langle 1  i  k+1  l+1\rangle + 
  w_1  x_2 \langle 1  i+1  k  l+1\rangle +  w_1  w_2 \langle 1  i+1  k+1  l+1\rangle 
   \nonumber \\&&+ x_2  y_1 \langle 1  k  l+1  j\rangle +  x_2  z_1 \langle 1  k  l+1  j+1\rangle + 
  w_2  y_1 \langle 1  k+1  l+1  j\rangle +  w_2  z_1 \langle 1  k+1  l+1  j+1\rangle   \nonumber\\&&
  + x_1  x_2  y_1 \langle i  k  l+1  j\rangle +  x_1  x_2  z_1 \langle i  k  l+1  j+1\rangle + 
  w_2  x_1  y_1 \langle i  k+1  l+1  j\rangle     \nonumber\\&&+ 
  w_2  x_1  z_1 \langle i  k+1  l+1  j+1\rangle  + 
  w_1  x_2  y_1 \langle i+1  k  l+1  j\rangle + 
  w_1  x_2  z_1 \langle i+1  k  l+1  j+1\rangle     \nonumber\\&&+ 
  w_1  w_2  y_1 \langle i+1  k+1  l+1  j\rangle+ 
  w_1  w_2  z_1 \langle i+1  k+1  l+1  j+1\rangle  \nonumber\\
  b&=&x_2 y_1  \langle 1  i+1  k  j \rangle  + x_2 z_1  \langle 1  i+1  k  j+1 \rangle  + 
 y_1 y_2  \langle 1  i+1  l  j \rangle  + y_2 z_1  \langle 1  i+1  l  j+1 \rangle     \nonumber\\&& + 
 y_1 z_2  \langle 1  i+1  l+1  j \rangle+ z_1 z_2  \langle 1  i+1  l+1  j+1 \rangle   \nonumber\\
 c&=&x_2 y_1 \langle 1  i  k  j\rangle  + x_2 z_1 \langle 1  i  k  j+1\rangle  + 
 w_2 y_1 \langle 1  i  k+1  j\rangle  + w_2 z_1 \langle 1  i  k+1  j+1\rangle  + 
 y_1 y_2 \langle 1  i  l  j\rangle    \nonumber\\&&+ y_2 z_1 \langle 1  i  l +1  j\rangle + 
 y_1 z_2 \langle 1  i  l+1  j\rangle  + z_1 z_2 \langle 1  i  l+1  j+1\rangle   \nonumber\\
 d&=&w_1 y_1 \langle 1  i+1  k+1  j\rangle  + w_1 z_1 \langle 1  i+1  k+1  j+1\rangle  \nonumber\\
 e&=& x_1  x_2  \langle 1  i  k  l \rangle  +  w_2  x_1  \langle 1  i  k+1  l \rangle  + 
  w_1  x_2  \langle 1  i+1  k  l \rangle  +  w_1  w_2  \langle 1  i+1  k+1  l \rangle  + 
  x_2  y_1  \langle 1  k  l  j \rangle    \nonumber\\&&+  x_2  z_1  \langle 1  k  l  j+1 \rangle  + 
  w_2  y_1  \langle 1  k+1  l  j \rangle  +  w_2  z_1  \langle 1  k+1  l  j+1 \rangle  + 
  x_1  x_2  y_1  \langle i  k  l  j \rangle     \nonumber\\&& +  x_1  x_2  z_1  \langle i  k  l  j+1 \rangle + 
  w_2  x_1  y_1  \langle i  k+1  l  j \rangle  +  w_2  x_1  z_1  \langle i  k+1  l  j+1 \rangle  + 
  w_1  x_2  y_1  \langle i+1  k  l  j \rangle      \nonumber\\&&+  w_1  x_2  z_1  \langle i+1  k  l  j+1 \rangle  + 
  w_1  w_2  y_1  \langle i+1  k+1  l  j \rangle  + 
  w_1  w_2  z_1  \langle i+1  k+1  l  j+1 \rangle .
\end{eqnarray}
We can calculate all forms for each pattern and the explicit form is written in the appendix. Then the full form of the 2-loop n-pt MHV log amplituhedron is 
\begin{equation}
\Omega[\log{[\mathcal{A}^{n\text{-pt 2-loop}}_{\text{MHV}}]}]=\sum_{\substack{i,j,k,l=2,3,\cdots,n-1\\i<k<l<j}}\Omega_{ijkl}^1[\log]+\sum_{i<k<j<l}\Omega_{ijkl}^2[\log]+\cdots+\sum_{k<l<i<j}\Omega_{ijkl}^{13}[\log].
\end{equation}
Then we can compare with this result and the non-planar sum of the double pentagon diagrams (\ref{eq:nonplanarsum}) and we checked that these results are corresponding up to at least 22-pt. \par
In the case of the 2-loop MHV amplitude, we can obtain the log of the amplitude from the canonical form on the well-defined space as \eqref{eq:loggeometry}.  However, the important point is that in the case of the 3-loop or higher loop, we can not define the log of the amplitude as a canonical form on the well-defined space. This means that the 2-loop MHV case is a special that we can define the log of the amplitude geometrically. 

\subsection{Square of the Amplituhedron and Positivity}
Next we consider the decomposition of the square of the 1-loop MHV amplituhedron. From (\ref{eq:square}), the square of the 1-loop MHV amplituhedron is decomposed into the amplituhedron and the log amplituhedron
\begin{eqnarray}
\big(\mathcal{L}_{i}=D_{i}\cdot Z \big)&=&\big( \mathcal{L}_{i}=D_{i}\cdot Z,\ \ \langle D_{(1)}D_{(2)} \rangle>0 \big)\nonumber\\&+& \big(\mathcal{L}_{i}=D_{i}\cdot Z,\ \ \langle D_{(1)}D_{(2)} \rangle<0\big)
\end{eqnarray}
for $i=1,2$. 
We can see this decomposition directly from the canonical form. For example, the 4-point case, the canonical form of the amplitude and log of the amplitude is 
\begin{eqnarray}
\label{eq:2loopcanonicalform}
\Omega[\mathcal{A}]=\frac{dx_1dx_2dw_1dw_2dy_1dy_2dz_1dz_2}{x_1x_2w_1w_2y_1y_2z_1z_2}\frac{x_1z_2+x_2z_1+w_1y_2+w_2y_1}{\{(x_1-x_2)(z_2-z_1)+(w_1-w_2)(y_2-y_1)\}}\nonumber\\
\Omega[\log{\mathcal{A}}]=\frac{dx_1dx_2dw_1dw_2dy_1dy_2dz_1dz_2}{x_1x_2w_1w_2y_1y_2z_1z_2}\frac{-(x_1z_1+x_2z_2+w_1y_1+w_2y_2)}{\{(x_1-x_2)(z_2-z_1)+(w_1-w_2)(y_2-y_1)\}}.\ \ \ 
\end{eqnarray}
Then
\begin{eqnarray}
\Omega[\mathcal{A}]+\Omega[\log{\mathcal{A}}]=\frac{dx_1dx_2dw_1dw_2dy_1dy_2dz_1dz_2}{x_1x_2w_1w_2y_1y_2z_1z_2}.
\end{eqnarray}
This is just the canonical form of the square of the 1-loop MHV amplituhedron (\ref{eq:square}). We can be confirmed that it holds for general $n$-point case from the explicit representation of the canonical form. \par
The interesting feature is that the numerator of the canonical form of the 2-loop MHV amplituhedron is the positive part of $\langle ABCD \rangle$ and the numerator of the log amplitude is the negative part. For example, the 4-pt case, 
\begin{eqnarray}
\langle ABCD \rangle =\langle 1234 \rangle\{x_1z_2+x_2z_1+w_1y_2+w_2y_1-(x_1z_1+x_2z_2+w_1y_1+w_2y_2)\}.
\end{eqnarray}
From the condition that $(A,B)$ and $(C,D)$ are the 1-loop MHV amplituhedron, we can see that
\begin{equation}
\langle 1234 \rangle, x_1,x_2,w_1,w_2,\cdots,z_2>0.
\end{equation}
Then $\langle ABCD \rangle$ is decomposed to 
\begin{equation}
\langle ABCD \rangle =A^++A^-
\end{equation}
where
\begin{equation}
A^+=\langle 1234 \rangle(x_1z_2+x_2z_1+w_1y_2+w_2y_1),\ \ \ A^-=-\langle 1234 \rangle(x_1z_1+x_2z_2+w_1y_1+w_2y_2)
\end{equation}
and $A^+$ is positive,  $A^-$ is negative. From (\ref{eq:2loopcanonicalform}),  
\begin{eqnarray}
\Omega[\mathcal{A}]=\frac{dx_1dx_2dw_1dw_2dy_1dy_2dz_1dz_2}{x_1x_2w_1w_2y_1y_2z_1z_2}\frac{A^+}{\langle ABCD \rangle}\\
\Omega[\log{\mathcal{A}}]=\frac{dx_1dx_2dw_1dw_2dy_1dy_2dz_1dz_2}{x_1x_2w_1w_2y_1y_2z_1z_2}\frac{A^-}{\langle ABCD \rangle},
\end{eqnarray}
we use the relation 
\begin{equation}
\langle 1234 \rangle\{(x_1-x_2)(z_2-z_1)+(w_1-w_2)(y_2-y_1)\}=\langle ABCD \rangle.
\end{equation}
From the $n$-point forms of the amplitude and the log amplitude, we can see that this holds for general $n$-point case. $\langle ABCD \rangle$ is decomposed into the positive and negative parts even for the $n$-pt case. For example, the pattern (1) for (\ref{eq:pattern}),
\begin{equation}
\langle ABCD \rangle = az_2-bw_1-cx_1-dw_2+ey_2
\end{equation}
where $a,b,c,d,e$ are defined as (\ref{eq:abcddef}) and these are positive. Then the positive and negative parts is
\begin{equation}
A^+= az_2+ey_2,\ \ \ A^-=-(bw_1+cx_1+dw_2).
\end{equation}
The canonical form of the 2-loop amplitude and the log of the amplitude for this pattern (1) is 
\begin{eqnarray}
\Omega_{ijkl}^1[\mathcal{A}]=\frac{dx_1dx_2dw_1dw_2dy_1dy_2dz_1dz_2}{x_1x_2w_1w_2y_1y_2z_1z_2}\frac{A^+}{\langle ABCD \rangle}\\
\Omega_{ijkl}^1[\log{\mathcal{A}}]=\frac{dx_1dx_2dw_1dw_2dy_1dy_2dz_1dz_2}{x_1x_2w_1w_2y_1y_2z_1z_2}\frac{A^-}{\langle ABCD \rangle}
\end{eqnarray}
and we can see that this holds for all another patterns of (\ref{eq:pattern}). From this result and $A^+>0$, the form of the $n$-pt 2-loop MHV amplituhedron is positive. Addition to this, in the form of the log amplitude, $A^-<0$ and $\langle ABCD \rangle<0$. Then the log of the amplitude is  also positive. The positivity of the canonical form is related to the existence of a "dual amplituhedron" \cite{positiveamplituhedron}. Then this is the another prove of the positivity of the canonical form directly.


\section{Conclusion}
In this work, we found the direct triangulation of the 2-loop MHV amplituhedron from the sign flip definition. These cells are non-redundant cells and live directly in $(AB), (CD)$ space. Unlike the BCFW case, there is no reference to any tree amplitudes.  We have seen that the $n$-point 2-loop MHV amplituhedron is triangulated into the $\frac{1}{4}(n-3)^2(n-2)^2$ cells and these are divided into 13 groups. These have spurious poles but in the sum of these cells, the spurious poles are canceled and only remain the physical poles. We checked these results are consistent with the local representation and the BCFW. \par
We obtained the $n$-point 2-loop MHV log amplitude from the geometry that constructed from the two 1-loop MHV amplituhedron and a negative constraint. However, this 2-loop MHV  case is a special that we can define the log of the amplitude geometrically. The 3-loop or higher loop case, we can not define the log of the amplitude as a canonical form on the well-defined space. \par
In the case of the 1-loop MHV amplituhedron, the triangulation from the sign flips corresponds to the BCFW triangulation. However, in the 2-loop MHV case, this triangulation doesn't correspond to the BCFW. Then the question is how to interpret this triangulation as the recursion relation. \par
The natural generalization is to go to the higher loop MHV amplituhedron. For example, the 3-loop MHV case, we have three 1-loop amplituhedron $(AB), (CD), (EF)$ and three constraints $\langle ABCD \rangle >0$, $\langle ABEF \rangle >0$, $\langle CDEF \rangle >0$.  In general $L$-loop MHV amplituhedron, there are $L$ 1-loop MHV amplituhedron and $\frac{1}{2}L(L-1)$ constraints. Is it possible to obtain the general canonical form of the 3 or general $L$-loop MHV amplituhedron from the sign flip triangulation?\par
Another generalization is to go to a higher $k$ amplituhedron. However, it is difficult to generalize this triangulation to the general $\text{N}^k$MHV amplituhedron. Because in the MHV case, we can use the isomorphism between the $k=2, m=2$ tree amplituhedron and the 1-loop MHV amplituhedron to obtain the direct triangulation from sign flips. However, there is no isomorphism between the $m=2$ amplituhedron and the $\text{N}^k$MHV loop amplituhedron. Then the future work is that to avoid using the isomorphism, we need to consider the direct relation between the triangulation of the $m=4$ amplituhedron and the sign flip patterns. \par
\section*{Acknowledgements}
We thank Jaroslav Trnka for first suggesting the problem, for numerous stimulating discussions and comments on the draft. This work was supported through the hospitality of the Center for Quantum Mathematics and Physics (QMAP), Department of Physics, University of California, Davis. This work is supported by the FY2017 Course-by-Course Education Program to Cultivate Researchers in Physical Sciences with Broad Perspectives, SOKENDAI.


\appendix
\section{Explicit Calculation of the 2-loop 5-point MHV Amplitude}
The case of $(2,3)\times(2,3)$,
\begin{eqnarray}
\begin{cases}
 Z_A=Z_1+x_1Z_2+w_1Z_3\\ Z_B=-Z_1+y_1Z_3+z_1Z_4\nonumber\\
 \end{cases}
 \begin{cases}
 Z_C=Z_1+x_2Z_2+w_2Z_3\\ Z_D=-Z_1+y_2Z_3+z_2Z_4
\end{cases}
\end{eqnarray} 
Therefore it is same as 4-pt case. $C$-matrix is (\ref{eq:4ptcmatrix}) and the form is (\ref{eq:4ptform}). \\
Next the case of $(3,4)\times(3,4)$, 
\begin{eqnarray}
\begin{cases}
 Z_A=Z_1+x_1Z_3+w_1Z_4\\ Z_B=-Z_1+y_1Z_4+z_1Z_5\nonumber\\
 \end{cases}
 \begin{cases}
 Z_C=Z_1+x_2Z_3+w_2Z_4\\ Z_D=-Z_1+y_2Z_4+z_2Z_5
\end{cases}
\end{eqnarray} 
Then
\begin{equation}
\langle ABCD\rangle=\langle 1345 \rangle \{(x_1-x_2)(y_1z_2-y_2z_1)+(z_1-z_2)(w_1x_2-w_2x_1)\}
\end{equation}
It is almost same as the case of $(2,3)\times(2,3)$. The only difference is $\langle 1345 \rangle $, thus there are 8 forms and the sum of these forms is
\begin{eqnarray}
\Omega_{3434}&=&\frac{dx_1dx_2dw_1dw_2dy_1dy_2dz_1dz_2}{x_1x_2w_1w_2y_1y_2z_1z_2}\frac{\langle 1345 \rangle}{\langle ABCD \rangle}(x_1y_1z_2+x_2y_2z_1+x_2w_1z_1+x_1w_2z_2)\nonumber\\
&=&\frac{\langle 1345 \rangle^3 \langle ABd^2A \rangle \langle ABd^2B\rangle \langle CDd^2C \rangle \langle CDd^2D\rangle}{\langle AB13 \rangle \langle AB15 \rangle\langle AB34 \rangle \langle AB45\rangle\langle ABCD \rangle\langle CD13 \rangle \langle CD15 \rangle\langle CD34 \rangle \langle CD45 \rangle}\nonumber\\
&\times&\biggr\{ \langle AB45 \rangle \langle CD13 \rangle+\langle AB34 \rangle \langle CD15 \rangle+\langle AB15 \rangle \langle CD34 \rangle+\langle AB13 \rangle \langle CD45 \rangle\biggl\}\ \ \ \ \ \ \ 
\end{eqnarray}
The case of $(2,3)\times(2,4)$, two 1-loop amplituhedron are parametrized as
\begin{eqnarray}
\begin{cases}
 Z_A=Z_1+x_1Z_2+w_1Z_3\\ Z_B=-Z_1+y_1Z_3+z_1Z_4\nonumber\\
 \end{cases}
 \begin{cases}
 Z_C=Z_1+x_2Z_2+w_2Z_3\\ Z_D=-Z_1+y_2Z_4+z_2Z_5
\end{cases}
\end{eqnarray}
In view of the $Y=C\cdot Z$ description, the $C$-matrix is
\begin{equation}
C = \left(
    \begin{array}{ccccc}
      1 & x_1 & w_1 & 0& 0  \\
      -1 &0 & y_1 & z_1  & 0 \\
      1 & x_2 & w_2 & 0 &0 \\
      -1 &0&0 & y_2 & z_2 
    \end{array}
  \right).
\end{equation}
The constraint is
\begin{eqnarray}
\langle ABCD\rangle&=&(x_1-x_2)\{y_1y_2\langle1234\rangle+y_1z_2\langle1235\rangle+z_1z_2\langle 1245\rangle\}\nonumber\\&+&(x_1w_2-x_2w_1)\{(y_2-z_1)\langle1234\rangle+z_2\langle1235\rangle-z_1z_2\langle2345\rangle\}\nonumber\\&+&(w_1-w_2)z_1z_2\langle1345\rangle
\end{eqnarray}
From $\langle ABCD \rangle>0$, 
\begin{eqnarray}
x_1&>&x_2-\frac{(x_1w_2-x_2w_1)\{(y_2-z_1)\langle1234\rangle+z_2\langle1235\rangle-z_1z_2\langle2345\rangle\}+(w_1-w_2)z_1z_2\langle1345\rangle}{y_1y_2\langle1234\rangle+y_1z_2\langle1235\rangle+z_1z_2\langle 1245\rangle}\nonumber\\
&=&x_2-a
\end{eqnarray}
The region of $x_1$ is depends on the sign of $a$. When $a<0$,  
\begin{eqnarray}
x_1>x_2-a,\ w_2<w_1-\frac{(x_1w_2-x_2w_1)\{(y_2-z_1)\langle1234\rangle+z_2\langle1235\rangle-z_1z_2\langle2345\rangle\}}{z_1z_2\langle1345\rangle}=w_1-b\nonumber\\
\end{eqnarray}
Similarly, the region of $w_1$ is depends on the sign of $b$. When $b<0$, there are 2 cases that
\begin{equation}
\begin{cases}
w_1<w_2-b,\ \ \text{and}\ \ x_2<\frac{w_2}{w_1}x_1,\ \ z_1>\frac{y_2 \langle 1234\rangle+z_2\langle 1235 \rangle}{\langle 1234 \rangle+z_2\langle 2345 \rangle}=c\\
 \ \ \ \ \ \ \ \ \ \ \ \ \text{or} \\
w_1<w_2-b,\ \ \text{and}\ \ x_2>\frac{w_2}{w_1}x_1,\ \ z_1<c
\end{cases}
\end{equation}
There are 8 cases depending on the signs of a,b. The forms for these cases can be obtained similarly as 4-point case,
\begin{eqnarray}
\Omega_1&=&\frac{1}{x_1-x_2+a}\left(\frac{1}{x_2}-\frac{1}{x_2-\frac{w_2}{w_1}x_1}\right)\left(\frac{1}{w_1}-\frac{1}{w_1-w_2+b}\right)\frac{1}{w_2}\frac{1}{y_1}\frac{1}{y_2}\frac{1}{z_1-c}\frac{1}{z_2}\nonumber\\
\Omega_2&=&\frac{1}{x_1-x_2+a}\frac{1}{x_2-\frac{w_2}{w_1}x_1}\left(\frac{1}{w_1}-\frac{1}{w_1-w_2+b}\right)\frac{1}{w_2}\frac{1}{y_1}\frac{1}{y_2}\left(\frac{1}{z_1}-\frac{1}{z_1-c}\right)\frac{1}{z_2}\nonumber\\
\Omega_3&=&\frac{1}{x_1-x_2+a}\left(\frac{1}{x_2}-\frac{1}{x_2-\frac{w_2}{w_1}x_1}\right)\frac{1}{w_1}\frac{1}{w_2-w_1-b}\frac{1}{y_1}\frac{1}{y_2}\left(\frac{1}{z_1}-\frac{1}{z_1-c}\right)\frac{1}{z_2}\nonumber\\
\Omega_4&=&\frac{1}{x_1-x_2+a}\frac{1}{x_2-\frac{w_2}{w_1}x_1}\frac{1}{w_1}\frac{1}{w_2-w_1-b}\frac{1}{y_1}\frac{1}{y_2}\frac{1}{z_1-c}\frac{1}{z_2}\nonumber\\
\Omega_5&=&\left(\frac{1}{x_2}-\frac{1}{x_2-x_1-a}\right) \frac{1}{x_1-\frac{w_1}{w_2}x_2}\frac{1}{w_1-w_2+b}\frac{1}{w_2}\frac{1}{y_1}\frac{1}{y_2}\frac{1}{z_1-c}\frac{1}{z_2}\\
\Omega_6&=&\left(\frac{1}{x_2}-\frac{1}{x_2-x_1-a}\right) \left(\frac{1}{x_1}-\frac{1}{x_1-\frac{w_1}{w_2}x_2}\right)\frac{1}{w_1-w_2+b}\frac{1}{w_2}\frac{1}{y_1}\frac{1}{y_2}\left(\frac{1}{z_1}-\frac{1}{z_1-c}\right)\frac{1}{z_2}\nonumber\\
\Omega_7&=&\left(\frac{1}{x_2}-\frac{1}{x_2-x_1-a}\right) \frac{1}{x_1-\frac{w_1}{w_2}x_2}\frac{1}{w_1}\left(\frac{1}{w_2}-\frac{1}{w_2-w_1-b}\right)\frac{1}{y_1}\frac{1}{y_2}\left(\frac{1}{z_1}-\frac{1}{z_1-c}\right)\frac{1}{z_2}\nonumber\\
\Omega_8&=&\left(\frac{1}{x_2}-\frac{1}{x_2-x_1-a}\right) \left(\frac{1}{x_1}-\frac{1}{x_1-\frac{w_1}{w_2}x_2}\right)\frac{1}{w_1}\left(\frac{1}{w_2}-\frac{1}{w_2-w_1-b}\right)\frac{1}{y_1}\frac{1}{y_2}\frac{1}{z_1-c}\frac{1}{z_2}\nonumber
\end{eqnarray}
For
\begin{eqnarray}
a&=&\frac{(x_1w_2-x_2w_1)\{(y_2-z_1)\langle1234\rangle+z_2\langle1235\rangle-z_1z_2\langle2345\rangle\}+(w_1-w_2)z_1z_2\langle1345\rangle}{y_1y_2\langle1234\rangle+y_1z_2\langle1235\rangle+z_1z_2\langle 1245\rangle}\nonumber\\
b&=&\frac{(x_1w_2-x_2w_1)\{(y_2-z_1)\langle1234\rangle+z_2\langle1235\rangle-z_1z_2\langle2345\rangle\}}{z_1z_2\langle1345\rangle}\nonumber\\
c&=&\frac{y_2 \langle 1234\rangle+z_2\langle 1235 \rangle}{\langle 1234 \rangle+z_2\langle 2345 \rangle}
\end{eqnarray}
Then sum of these 8 forms is
\begin{eqnarray}
\Omega_{2324}&=&\frac{dx_1dx_2dw_1dw_2dy_1dy_2dz_1dz_2}{x_1x_2w_1w_2y_1y_2z_1z_2}\frac{1}{\langle ABCD \rangle}\left\{\langle 1234 \rangle (x_1w_2y_2+x_1y_1y_2+x_2w_1z_1)\right.\nonumber\\
&+&\left.\langle 1235 \rangle x_1z_2(w_2+y_1)+z_1z_2(\langle 1345 \rangle w_1 +\langle 1245 \rangle x_1+\langle 2345 \rangle x_2w_1)\right\}
\end{eqnarray}
Rewrite it into the momentum twistor,
\begin{eqnarray}
\Omega_{2324}&=&\frac{\langle 123A_3 \rangle \langle 123C_4 \rangle\langle ABd^2A \rangle \langle ABd^2B\rangle \langle CDd^2C \rangle \langle CDd^2D\rangle}{\left\{\LARGE{\substack{\langle AB12 \rangle\langle AB13 \rangle^2\langle AB14\rangle\langle AB23\rangle\langle AB34\rangle\\\times\langle ABCD \rangle\langle CD12 \rangle\langle CD13\rangle\langle CD14\rangle\langle CD15\rangle\langle CD23\rangle\langle CD45\rangle}}\right\}}\nonumber
\\
&\times&\biggl\{\langle 123 A_3\rangle(\langle AB12 \rangle\langle CD13 \rangle\langle CD45 \rangle+\langle AB15 \rangle\langle CD14 \rangle\langle CD23 \rangle)\nonumber\\
&&\ \ \langle 123 C_4\rangle(\langle AB13 \rangle\langle AB34 \rangle\langle CD12 \rangle+\langle AB13 \rangle\langle AB14 \rangle\langle CD23 \rangle)\nonumber\\
&&\ \ -\langle 1235\rangle\langle AB13 \rangle\langle AB14 \rangle\langle CD14 \rangle\langle CD23 \rangle-\langle 2345\rangle\langle AB12 \rangle\langle AB13 \rangle\langle CD12 \rangle\langle CD14 \rangle\biggr\}\nonumber.\\
\end{eqnarray}
We use these symbols
\begin{eqnarray}
(AB)\cap(1ii+1)=-Z_1\langle ii+1AB \rangle-Z_i\langle i+1AB1 \rangle-Z_{i+1}\langle AB 1i \rangle &\equiv& A_i\nonumber\\
(CD)\cap(1ii+1)=-Z_1\langle ii+1CD \rangle-Z_i\langle i+1CD1 \rangle-Z_{i+1}\langle CD 1i \rangle &\equiv& C_i
\end{eqnarray}
The case of $(2,4)\times(3,4)$, 
\begin{eqnarray}
\begin{cases}
 Z_A=Z_1+x_1Z_2+w_1Z_3\\ Z_B=-Z_1+y_1Z_4+z_1Z_5\nonumber\\
 \end{cases}
 \begin{cases}
 Z_C=Z_1+x_2Z_3+w_2Z_4\\ Z_D=-Z_1+y_2Z_4+z_2Z_5
\end{cases}
\end{eqnarray}
$C$-matrix is
\begin{equation}
C = \left(
    \begin{array}{ccccc}
      1 & x_1 & w_1 & 0& 0  \\
      -1 &0 &0& y_1 & z_1   \\
      1 &0& x_2 & w_2 &0 \\
      -1 &0&0 & y_2 & z_2 
    \end{array}
  \right)
\end{equation}
\begin{eqnarray}
\langle ABCD\rangle&=&(z_2-z_1)(\langle 1345 \rangle w_1w_2+\langle 1235 \rangle x_1x_2+\langle 1245 \rangle x_1w_2)\nonumber\\
&+&(z_1y_2-z_2y_1)\{\langle 1345 \rangle(x_2-w_1)-\langle 1245 \rangle x_1+ \langle 2345 \rangle x_1x_2 \}\nonumber\\
&+&(y_2-y_1)\langle 1234 \rangle x_1x_2
\end{eqnarray}
From$\langle ABCD \rangle>0$, 
\begin{eqnarray}
z_2&>&z_1-\frac{(z_1y_2-z_2y_1)\{\langle 1345\rangle (x_2-w_1)-\langle 1245 \rangle x_1+ \langle 2345 \rangle x_1x_2 \}
+(y_2-y_1)\langle 1234 \rangle x_1x_2}{x_1x_2 \langle 1234 \rangle}\nonumber\\
&=&z_1-a
\end{eqnarray}
The region of $z_2$ is depends on the sign of $a$. When $a<0$,  
\begin{equation}
z_2>z_1-a,\ \text{and}\ \ y_2<y_1-\frac{(z_1y_2-z_2y_1)\{\langle 1345\rangle (x_2-w_1)-\langle 1245 \rangle x_1+ \langle 2345 \rangle x_1x_2 \}}{x_1x_2 \langle 1234 \rangle}=y_1-b
\end{equation}
Similarly, the region of $y_2$ is depends on the sign of $b$. When $b<0$, there are 2 cases that
\begin{equation}
y_2<y_1-b,\ \ \text{and}\ \ \begin{cases}
 z_1>\frac{y_1}{y_2}z_2,\ \ x_2<\frac{w_1 \langle 1345\rangle+x_1\langle 1245 \rangle}{x_1\langle 2345 \rangle+\langle 1345 \rangle}=c\\\ \ \ \ \ \  \text{or} \\
z_1<\frac{y_1}{y_2}z_2,\ \ x_2>c
\end{cases}
\end{equation}
There are 8 cases depending on the signs of a,b. Then the forms for these cases are
\begin{eqnarray}
\Omega_1&=&\frac{1}{x_1}\left(\frac{1}{x_2}-\frac{1}{x_2-c}\right)\frac{1}{w_1}\frac{1}{w_2}\frac{1}{y_1}\left(\frac{1}{y_2}-\frac{1}{y_2-y_1+b}\right)\frac{1}{z_1-\frac{y_1}{y_2}z_2}\frac{1}{z_2-z_1+a}\nonumber\\
\Omega_2&=&\frac{1}{x_1}\frac{1}{x_2-c}\frac{1}{w_1}\frac{1}{w_2}\frac{1}{y_1}\left(\frac{1}{y_2}-\frac{1}{y_2-y_1+b}\right)\left(\frac{1}{z_1}-\frac{1}{z_1-\frac{y_1}{y_2}z_2}\right)\frac{1}{z_2-z_1+a}\nonumber\\
\Omega_3&=&\frac{1}{x_1}\frac{1}{x_2-c}\frac{1}{w_1}\frac{1}{w_2}\frac{1}{y_1-y_2-b}\frac{1}{y_2}\frac{1}{z_1-\frac{y_1}{y_2}z_2}\frac{1}{z_2-z_1+a}\nonumber\\
\Omega_4&=&\frac{1}{x_1}\left(\frac{1}{x_2}-\frac{1}{x_2-c}\right)\frac{1}{w_1}\frac{1}{w_2}\frac{1}{y_1}\frac{1}{y_2-y_1+b}\left(\frac{1}{z_1}-\frac{1}{z_1-\frac{y_1}{y_2}z_2}\right)\frac{1}{z_2-z_1+a}\nonumber\\
\Omega_5&=&\frac{1}{x_1}\left(\frac{1}{x_2}-\frac{1}{x_2-c}\right)\frac{1}{w_1}\frac{1}{w_2}\frac{1}{y_1}\frac{1}{y_2-y_1+b}\left(\frac{1}{z_1}-\frac{1}{z_1-\frac{y_1}{y_2}z_2}\right)\left(\frac{1}{z_2}-\frac{1}{z_2-z_1+a}\right)\nonumber\\
\Omega_6&=&\frac{1}{x_1}\frac{1}{x_2-c}\frac{1}{w_1}\frac{1}{w_2}\frac{1}{y_1}\frac{1}{y_2-y_1+b}\left(\frac{1}{z_1}-\frac{1}{z_1-\frac{y_1}{y_2}z_2}\right)\frac{1}{z_2-z_1+a}\\
\Omega_7&=&\frac{1}{x_1}\frac{1}{x_2-c}\frac{1}{w_1}\frac{1}{w_2}\left(\frac{1}{y_1}-\frac{1}{y_1-y_2-b}\right)\frac{1}{y_2}\left(\frac{1}{z_1}-\frac{1}{z_1-\frac{y_1}{y_2}z_2}\right)\left(\frac{1}{z_2}-\frac{1}{z_2-z_1+a}\right)\nonumber\\
\Omega_8&=&\frac{1}{x_1}\left(\frac{1}{x_2}-\frac{1}{x_2-c}\right)\frac{1}{w_1}\frac{1}{w_2}\left(\frac{1}{y_1}-\frac{1}{y_1-y_2-b}\right)\frac{1}{y_2}\left(\frac{1}{z_1}-\frac{1}{z_1-\frac{y_1}{y_2}z_2}\right)\frac{1}{z_2-z_1+a}\nonumber
\end{eqnarray}
For
\begin{eqnarray}
a&=&\frac{(z_1y_2-z_2y_1)\{\langle 1345\rangle (x_2-w_1)-\langle 1245 \rangle x_1+ \langle 2345 \rangle x_1x_2 \}
+(y_2-y_1)\langle 1234 \rangle x_1x_2}{x_1x_2 \langle 1234 \rangle}\nonumber\\
b&=&\frac{(z_1y_2-z_2y_1)\{\langle 1345\rangle (x_2-w_1)-\langle 1245 \rangle x_1+ \langle 2345 \rangle x_1x_2 \}}{x_1x_2 \langle 1234 \rangle}\nonumber\\
c&=&\frac{w_1 \langle 1345\rangle+x_1\langle 1245 \rangle}{x_1\langle 2345 \rangle+\langle 1345 \rangle}
\end{eqnarray}
Then sum of these 8 forms is
\begin{eqnarray}
\Omega_{2434}&=&\frac{dx_1dx_2dw_1dw_2dy_1dy_2dz_1dz_2}{x_1x_2w_1w_2y_1y_2z_1z_2}\frac{1}{\langle ABCD \rangle}\left\{\langle 1345 \rangle (w_1w_2z_2+w_1y_1z_2+x_2y_2z_1)\right.\nonumber\\
&+&\left.\langle 1235 \rangle x_1x_2z_2+\langle 1245 \rangle (x_1w_2z_2+x_1y_1z_2)+\langle 1234 \rangle x_1x_2y_2 +\langle 2345 \rangle x_1x_2y_2z_1\right\}\nonumber\\
\end{eqnarray}
In the momentum twistor,
\begin{eqnarray}
\Omega_{2434}&=&\frac{\langle 123A_4\rangle\langle 134C_4 \rangle \langle ABd^2A \rangle \langle ABd^2B\rangle \langle CDd^2C \rangle \langle CDd^2D\rangle}{\left\{\LARGE{\substack{\langle AB12 \rangle\langle AB13\rangle\langle AB14 \rangle\langle AB15 \rangle  \langle AB23\rangle\langle AB45\rangle\\\times\langle ABCD \rangle\langle CD13 \rangle\langle CD14\rangle^2\langle CD15\rangle\langle CD34\rangle\langle CD45\rangle}}\right\}}\nonumber
\\
&\times&\biggl\{\langle 123 A_4\rangle(\langle AB45 \rangle\langle CD13 \rangle\langle CD14\rangle+\langle AB15 \rangle\langle CD34 \rangle\langle CD14\rangle)\\
&&\ \ -\langle 345 A_2\rangle\langle AB14 \rangle\langle CD14\rangle\langle CD15 \rangle+\langle 123C_4\rangle\langle CD14\rangle\langle AB45 \rangle\langle AB13 \rangle\biggr\}\nonumber
\end{eqnarray}
\\
The case of $(2,3)\times(3,4)$, 
\begin{eqnarray}
\begin{cases}
 Z_A=Z_1+x_1Z_2+w_1Z_3\\ Z_B=-Z_1+y_1Z_3+z_1Z_4\nonumber\\
 \end{cases}
 \begin{cases}
 Z_C=Z_1+x_2Z_3+w_2Z_4\\ Z_D=-Z_1+y_2Z_4+z_2Z_5
\end{cases}
\end{eqnarray}
$C$-matrix is
\begin{equation}
C = \left(
    \begin{array}{ccccc}
      1 & x_1 & w_1 & 0& 0  \\
      -1 &0 & y_1 & z_1  & 0 \\
      1 &0& x_2 & w_2  &0 \\
      -1 &0&0 & y_2 & z_2 
    \end{array}
  \right)
\end{equation}
\begin{eqnarray}
\langle ABCD\rangle&=&(y_1w_2-z_1x_2)(\langle 1345 \rangle z_2+\langle 1234 \rangle x_1+\langle 2345 \rangle x_1z_2)\nonumber\\
&+&z_2(z_1+w_2)(\langle 1345 \rangle w_1+\langle 1245 \rangle x_1)+x_1(y_1+x_2)( \langle 1235 \rangle z_2+\langle 1234 \rangle y_2 )\nonumber\\
\end{eqnarray}
In this case, from $\langle ABCD \rangle >0$,
\begin{equation}
y_1w_2-z_1x_2>-\frac{z_2(z_1+w_2)(\langle 1345 \rangle w_1+\langle 1245 \rangle x_1)+x_1(y_1+x_2)( \langle 1235 \rangle z_2+\langle 1234 \rangle y_2 )}{\langle 1345 \rangle z_2+\langle 1234 \rangle x_1+\langle 2345 \rangle x_1z_2}=-a
\end{equation}
However, from $x_1,x_2,w_1,w_2,y_1,y_2,z_1,z_2>0$, $a>0$. Therefore
\begin{equation}
z_1<\frac{w_2y_1}{x_2}+\frac{a}{x_2}
\end{equation}
Then the form is
\begin{eqnarray}
\Omega_{2334}&=&\frac{dx_1dx_2dw_1dw_2dy_1dy_2dz_1dz_2}{x_1x_2w_1w_2y_1y_2z_2}\left(\frac{1}{z_1}-\frac{1}{z_1-\frac{w_2y_1}{x_2}-\frac{a}{x_2}}\right)\nonumber\\
&=&\frac{dx_1dx_2dw_1dw_2dy_1dy_2dz_1dz_2}{x_1x_2w_1w_2y_1y_2z_1z_2}\frac{1}{\langle ABCD \rangle}\{w_2y_1( \langle 1345 \rangle z_2 +\langle 1234 \rangle x_1 +\langle 2345 \rangle x_1z_2)\nonumber\\
&+&z_2(z_1+w_2)(\langle 1345 \rangle w_1 +\langle 1245 \rangle x_1)+x_1(y_1+x_2)(\langle 1235 \rangle z_2 +\langle 1234 \rangle y_2)\}
\end{eqnarray}\\
We can write it in the momentum twistor space,
\begin{eqnarray}
\Omega_{2334}&=&\frac{\langle 123A_3 \rangle \langle 134C_4\rangle\langle ABd^2A \rangle \langle ABd^2B\rangle \langle CDd^2C \rangle \langle CDd^2D\rangle}{\left\{\LARGE{\substack{\langle AB12 \rangle\langle AB13 \rangle^2\langle AB14\rangle\langle AB23\rangle\langle AB34\rangle\\\times\langle ABCD \rangle\langle CD13 \rangle\langle CD14\rangle^2\langle CD15\rangle\langle CD34\rangle\langle CD45\rangle}}\right\}}\nonumber
\\
&\times&\biggl\{\langle AB13 \rangle\langle 123 C_4\rangle\langle CD4 A_3\rangle-\langle AB13 \rangle\langle AB14\rangle\langle CD13\rangle\langle 234 C_4\rangle\nonumber\ \ \ \ \ \ \ \\
&&\ \ +\langle CD14 \rangle\langle 145 A_2\rangle\langle CD3 A_3\rangle-\langle AB14 \rangle\langle AB23\rangle\langle CD13\rangle\langle CD14\rangle\langle1345 \rangle\biggr\}\ \ \ \ 
\end{eqnarray}
The case of $(2,4)\times(2,4)$, 
\begin{eqnarray}
\begin{cases}
 Z_A=Z_1+x_1Z_2+w_1Z_3\\ Z_B=-Z_1+y_1Z_4+z_1Z_5\nonumber\\
 \end{cases}
 \begin{cases}
 Z_C=Z_1+x_2Z_2+w_2Z_3\\ Z_D=-Z_1+y_2Z_4+z_2Z_5
\end{cases}
\end{eqnarray}
$C$-matrix is
\begin{equation}
C = \left(
    \begin{array}{ccccc}
      1 & x_1 & w_1 & 0& 0  \\
      -1 &0&0 & y_1 & z_1   \\
      1 & x_2 & w_2 & 0 &0 \\
      -1 &0&0 & y_2 & z_2 
    \end{array}
  \right)
\end{equation}
\begin{eqnarray}
\langle ABCD\rangle&=&(y_2-y_1)(x_1w_2-x_2w_1)\langle 1234 \rangle +(z_2-z_1)(x_1w_2-x_2w_1)\langle 1235 \rangle \nonumber\\
&+&(z_1y_2-z_2y_1)\{\langle 1245 \rangle (x_2-x_1)+\langle 1345 \rangle (w_2-w_1)+\langle 2345 \rangle (x_1w_2-x_2w_1)\}\nonumber\\
\end{eqnarray}
When $(x_1w_2-x_2w_1)>0,\ (z_1y_2-z_2y_1)>0$, from $\langle ABCD \rangle$,
\begin{equation}
y_2>y_1-a
\end{equation}
For
\begin{eqnarray}
a&=&\frac{(z_2-z_1)(x_1w_2-x_2w_1)\langle 1235 \rangle  }{(x_1w_2-x_2w_1)\langle 1234 \rangle}\nonumber\\
&+&\frac{(z_1y_2-z_2y_1)\{\langle 1245 \rangle (x_2-x_1)+\langle 1345 \rangle (w_2-w_1)+\langle 2345 \rangle (x_1w_2-x_2w_1)  }{(x_1w_2-x_2w_1)\langle 1234 \rangle}\ \ \ 
\end{eqnarray}
The region of $y_2$ is depends on the sign of $a$. When $a<0$,
\begin{eqnarray}
y_2>y_1-a\ ,\ z_2&<&z_1-\frac{(z_1y_2-z_2y_1)\{\langle 1245 \rangle (x_2-x_1)+\langle 1345 \rangle (w_2-w_1)+\langle 2345 \rangle (x_1w_2-x_2w_1) \} }{(x_1w_2-x_2w_1)\langle 1235 \rangle}\nonumber\\
&=&z_1-b
\end{eqnarray}
Similarly, the region of $z_2$ is depends on the sign of $b$. When $b<0$, 
\begin{equation}
z_2<z_1-b\ \text{and}\ \ x_2<x_1-\frac{\langle 1345 \rangle (w_2-w_1)+\langle 2345 \rangle (x_1w_2-x_2w_1) }{\langle 1245 \rangle}=x_1-c
\end{equation}
When $c<0$, 
\begin{equation}
x_2<x_1-c\ \text{and}\ \ w_2<w_1-\frac{\langle 2345 \rangle (x_1w_2-x_2w_1) }{\langle 1345 \rangle}=w_1-d
\end{equation}
From $w_2>0,  (x_1w_2-x_2w_1)>0,$ then $d>0$ and there are 8 cases depending on the signs of a,b,c.
\begin{eqnarray}
\Omega_1&=&\frac{1}{x_1-\frac{w_1}{w_2}x_2}\left(\frac{1}{x_2}-\frac{1}{x_2-x_1+c}\right)\frac{1}{w_1-w_2-d}\frac{1}{w_2}\frac{1}{y_1}\frac{1}{y_2-y_1+a}\frac{1}{z_1-\frac{y_1}{y_2}z_2}\left(\frac{1}{z_2}-\frac{1}{z_2-z_1+c}\right)\nonumber\\
\Omega_2&=&\frac{1}{x_1-x_2-c}\left(\frac{1}{x_2}-\frac{1}{x_2-\frac{w_2}{w_1}x_1}\right)\left(\frac{1}{w_1}-\frac{1}{w_1-w_2-d}\right)\frac{1}{w_2}\frac{1}{y_1}\frac{1}{y_2-y_1+a}\nonumber\\
&\times&\frac{1}{z_1-\frac{y_1}{y_2}z_2}\left(\frac{1}{z_2}-\frac{1}{z_2-z_1+b}\right)\nonumber\\
\Omega_3&=&\frac{1}{x_1-\frac{w_1}{w_2}x_2}\frac{1}{x_2-x_1+c}\frac{1}{w_1-w_2-d}\frac{1}{w_2}\frac{1}{y_1}\frac{1}{y_2-y_1+a}\frac{1}{z_1-z_2-b}\left(\frac{1}{z_2}-\frac{1}{z_2-\frac{y_2}{y_1}z_1}\right)\nonumber\\
\Omega_4&=&\left(\frac{1}{x_1}-\frac{1}{x_1-x_2-c}\right)\left(\frac{1}{x_2}-\frac{1}{x_2-\frac{w_2}{w_1}x_1}\right)\left(\frac{1}{w_1}-\frac{1}{w_1-w_2-d}\right)\frac{1}{w_2}\frac{1}{y_1}\frac{1}{y_2-y_1+a}\nonumber\\
&\times&\frac{1}{z_1-z_2-b}\left(\frac{1}{z_2}-\frac{1}{z_2-\frac{y_2}{y_1}z_1}\right)\nonumber\\
\Omega_5&=&\frac{1}{x_1-\frac{w_1}{w_2}x_2}\left(\frac{1}{x_2}-\frac{1}{x_2-x_1+c}\right)\frac{1}{w_1-w_2-d}\frac{1}{w_2}\left(\frac{1}{y_1}-\frac{1}{y_1-y_2-a}\right)\frac{1}{y_2}\frac{1}{z_1-\frac{y_1}{y_2}z_2}\frac{1}{z_2-z_1+b}\nonumber\\
\Omega_6&=&\frac{1}{x_1-x_2-c}\left(\frac{1}{x_2}-\frac{1}{x_2-\frac{w_2}{w_1}x_1}\right)\left(\frac{1}{w_1}-\frac{1}{w_1-w_2-d}\right)\frac{1}{w_2}\left(\frac{1}{y_1}-\frac{1}{y_1-y_2-a}\right)\frac{1}{y_2}\nonumber\\
&\times&\frac{1}{z_1-\frac{y_1}{y_2}z_2}\frac{1}{z_2-z_1+b}\nonumber\\
\Omega_7&=&\frac{1}{x_1-\frac{w_1}{w_2}x_2}\frac{1}{x_2-x_1+c}\frac{1}{w_1-w_2-d}\frac{1}{w_2}\left(\frac{1}{y_1}-\frac{1}{y_1-y_2-a}\right)\frac{1}{y_2}\nonumber\\
&\times&\left(\frac{1}{z_1}-\frac{1}{z_1-z_2-b}\right)\left(\frac{1}{z_2}-\frac{1}{z_2-\frac{y_2}{y_1}z_1}\right)\nonumber\\
\Omega_8&=&\left(\frac{1}{x_1}-\frac{1}{x_1-x_2-c}\right)\left(\frac{1}{x_2}-\frac{1}{x_2-\frac{w_2}{w_1}x_1}\right)\left(\frac{1}{w_1}-\frac{1}{w_1-w_2-d}\right)\frac{1}{w_2}\left(\frac{1}{y_1}-\frac{1}{y_1-y_2-a}\right)\frac{1}{y_2}\nonumber\\
&\times&\left(\frac{1}{z_1}-\frac{1}{z_1-z_2-b}\right)\left(\frac{1}{z_2}-\frac{1}{z_2-\frac{y_2}{y_1}z_1}\right)
\end{eqnarray}
For
\begin{eqnarray}
a&=&\frac{(z_2-z_1)(x_1w_2-x_2w_1)\langle 1235 \rangle  }{(x_1w_2-x_2w_1)\langle 1234 \rangle}\nonumber\\
&+&\frac{(z_1y_2-z_2y_1)\{\langle 1245 \rangle (x_2-x_1)+\langle 1345 \rangle (w_2-w_1)+\langle 2345 \rangle (x_1w_2-x_2w_1)  }{(x_1w_2-x_2w_1)\langle 1234 \rangle}\nonumber\\
b&=&\frac{(z_1y_2-z_2y_1)\{\langle 1245 \rangle (x_2-x_1)+\langle 1345 \rangle (w_2-w_1)+\langle 2345 \rangle (x_1w_2-x_2w_1) \} }{(x_1w_2-x_2w_1)\langle 1235 \rangle}\nonumber\\
c&=&\frac{\langle 1345 \rangle (w_2-w_1)+\langle 2345 \rangle (x_1w_2-x_2w_1) }{\langle 1245 \rangle}\nonumber\\
d&=&\frac{\langle 2345 \rangle (x_1w_2-x_2w_1) }{\langle 1345 \rangle}
\end{eqnarray}
This is the case of $(x_1w_2-x_2w_1)>0,\ (z_1y_2-z_2y_1)>0$. Next we consider the case of $(x_1w_2-x_2w_1)>0,\ (z_1y_2-z_2y_1)<0$. Forms are obtained by replacement as follows.
\begin{eqnarray}
\left(\frac{1}{x_1}-\frac{1}{x_1-x_2-c}\right) &\leftrightarrow& \frac{1}{x_1-x_2-c}\nonumber\\
\left(\frac{1}{x_2}-\frac{1}{x_2-x_1+c}\right) &\leftrightarrow& \frac{1}{x_2-x_1+c}\nonumber\\
\frac{1}{z_1-\frac{y_1}{y_2}z_2} &\to& \left(\frac{1}{z_1}-\frac{1}{z_1-\frac{y_1}{y_2}z_2}\right)\nonumber\\
 \left(\frac{1}{z_2}-\frac{1}{z_2-\frac{y_2}{y_1}z_1}\right) &\to& \frac{1}{z_2-\frac{y_2}{y_1}z_1}
\end{eqnarray}
Then the cases of $(x_1w_2-x_2w_1)<0,\ (z_1y_2-z_2y_1)<0$ and  $(x_1w_2-x_2w_1)<0,\ (z_1y_2-z_2y_1)>0$ are obtained that swap $1\leftrightarrow 2$ for the case of  $(x_1w_2-x_2w_1)>0,\ (z_1y_2-z_2y_1)>0$ and  $(x_1w_2-x_2w_1)>0,\ (z_1y_2-z_2y_1)<0$. 
Then sum of these 32 forms is\\
\begin{eqnarray}
\!\!\!\!\!\!\!\!\Omega_{2424}&=&\frac{dx_1dx_2\cdots dz_1dz_2}{x_1x_2w_1w_2y_1y_2z_1z_2}\frac{1}{\langle ABCD \rangle}\nonumber\\
&\times&\{\langle 1234 \rangle (x_2w_1y_1+x_1w_2y_2)+y_2z_1(\langle 1345 \rangle w_2+\langle 1245 \rangle x_2+\langle 2345 \rangle x_1w_2)\nonumber\\
&+&y_1z_2(\langle 1345 \rangle w_1+\langle 1245 \rangle x_1+\langle 2345 \rangle x_2w_1)+\langle 1235 \rangle (x_2w_1z_1+x_1w_2z_2)\}.
\end{eqnarray}
In the momentum twistor space,
\begin{eqnarray}
\Omega_{2424}&=&\frac{\langle 123A_4 \rangle \langle 123C_4\rangle\langle ABd^2A \rangle \langle ABd^2B\rangle \langle CDd^2C \rangle \langle CDd^2D\rangle}{\left\{\LARGE{\substack{\langle AB12 \rangle\langle AB13\rangle\langle AB14\rangle\langle AB15\rangle\langle AB23\rangle\langle AB45\rangle\langle ABCD \rangle\\\times\langle CD12 \rangle\langle CD13\rangle\langle CD14\rangle\langle CD15\rangle\langle CD23\rangle\langle CD45\rangle}}\right\}}\nonumber
\\
&\times&\biggl\{\langle 123 A_4\rangle(\langle AB12 \rangle\langle CD13 \rangle\langle CD45\rangle+\langle AB15 \rangle\langle CD14 \rangle\langle CD23 \rangle)\\
&&+\langle 123 C_4\rangle(\langle AB13 \rangle\langle AB45 \rangle\langle CD12 \rangle+\langle AB14 \rangle\langle AB23 \rangle\langle CD15 \rangle)\nonumber\\
&&+\langle 2345 \rangle(\langle AB12 \rangle\langle AB15 \rangle\langle CD13 \rangle\langle CD14 \rangle+\langle AB13 \rangle\langle AB14 \rangle\langle CD12 \rangle\langle CD15 \rangle)\biggr\}\nonumber
\end{eqnarray} 
The remaining patterns are $(3,4)\times(2,3), (2,4)\times(2,3),(3,4)\times(2,4)$. These forms can be obtained from $\Omega_{2334},\Omega_{2324},\Omega_{2434}$ that swap $AB \leftrightarrow CD$. 
\section{Explicit Results of the 2-loop n-point MHV Amplituhedron}
First we consider the $(1)$ case $i<k<l<j$,
\begin{eqnarray*}
&&\langle ABCD \rangle =\\
&& x_1 x_2 y_2 \langle 1  i  k  l\rangle  +  x_1 x_2 z_2 \langle 1  i  k  l+1\rangle  + 
 w_2  x_1 y_2 \langle 1  i  k+1  l\rangle  + w_2  x_1 z_2 \langle 1  i  k+1  l+1\rangle  - 
  x_1 (x_2 y_1 \langle 1  i  k  j\rangle  \\&&+ x_2 z_1 \langle 1  i  k  j+1\rangle  + 
    w_2 y_1 \langle 1  i  k+1  j\rangle  + w_2 z_1 \langle 1  i  k+1  j+1\rangle  + 
    y_1 y_2 \langle 1  i  l  j\rangle  + y_2 z_1 \langle 1  i  l  j+1\rangle  \\&&+ 
    y_1 z_2 \langle 1  i  l+1  j\rangle  + z_1 z_2 \langle 1  i  l+1  j+1\rangle ) + 
 w_1 x_2 y_2 \langle 1  i+1  k  l\rangle  + w_1 x_2 z_2 \langle 1  i+1  k  l+1\rangle \\&& - 
 w_2 (w_1 y_1 \langle 1  i+1  k+1  j\rangle  + 
    w_1 z_1 \langle 1  i+1  k+1  j+1\rangle ) + 
 w_1 w_2 y_2 \langle 1  i+1  k+1  l\rangle  + 
 w_1 w_2 z_2 \langle 1  i+1  k+1  l+1\rangle\\&&  - 
 w_1 (x_2 y_1 \langle 1  i+1  k  j\rangle  + x_2 z_1 \langle 1  i+1  k  j+1\rangle  + 
    y_1 y_2 \langle 1  i+1  l  j\rangle  + y_2 z_1 \langle 1  i+1  l  j+1\rangle  + 
    y_1 z_2 \langle 1  i+1  l+1  j\rangle \\&& + 
    z_1 z_2 \langle 1  i+1  l+1  j+1\rangle ) + x_2 y_1 y_2 \langle 1  k  l  j\rangle  + 
 x_2 y_2 z_1 \langle 1  k  l  j+1\rangle  + x_2 y_1 z_2 \langle 1  k  l+1  j\rangle  + 
 x_2 z_1 z_2 \langle 1  k  l+1  j+1\rangle  \\&&+ w_2 y_1 y_2 \langle 1  k+1  l  j\rangle  + 
 w_2 y_2 z_1 \langle 1  k+1  l  j+1\rangle  + 
 w_2 y_1 z_2 \langle 1  k+1  l+1  j\rangle  + 
 w_2 z_1 z_2 \langle 1  k+1  l+1  j+1\rangle  \\&&+ 
  x_1 x_2 y_1 y_2 \langle i  k  l  j\rangle  +  x_1 x_2 y_2 z_1 \langle i  k  l  j+1\rangle  + 
  x_1 x_2 y_1 z_2 \langle i  k  l+1  j\rangle  + 
  x_1 x_2 z_1 z_2 \langle i  k  l+1  j+1\rangle  \\&&+ 
 w_2  x_1 y_1 y_2 \langle i  k+1  l  j\rangle  + 
 w_2  x_1 y_2 z_1 \langle i  k+1  l  j+1\rangle  + 
 w_2  x_1 y_1 z_2 \langle i  k+1  l+1  j\rangle  + 
 w_2  x_1 z_1 z_2 \langle i  k+1  l+1  j+1\rangle \\&& + 
 w_1 x_2 y_1 y_2 \langle i+1  k  l  j\rangle  + 
 w_1 x_2 y_2 z_1 \langle i+1  k  l  j+1\rangle  + 
 w_1 x_2 y_1 z_2 \langle i+1  k  l+1  j\rangle  + 
 w_1 x_2 z_1 z_2 \langle i+1  k  l+1  j+1\rangle  \\&&+ 
 w_1 w_2 y_1 y_2 \langle i+1  k+1  l  j\rangle  + 
 w_1 w_2 y_2 z_1 \langle i+1  k+1  l  j+1\rangle  + 
 w_1 w_2 y_1 z_2 \langle i+1  k+1  l+1  j\rangle  \\&&+ 
 w_1 w_2 z_1 z_2 \langle i+1  k+1  l+1  j+1\rangle \\
 &&=az_2-bw_1-cx_1-dw_2+ey_2
\end{eqnarray*}
for
\begin{eqnarray*}
\label{eq:abcde}
a&= &x_1  x_2 \langle 1  i  k  l+1\rangle +  w_2  x_1 \langle 1  i  k+1  l+1\rangle + 
  w_1  x_2 \langle 1  i+1  k  l+1\rangle +  w_1  w_2 \langle 1  i+1  k+1  l+1\rangle +
   x_2  y_1 \langle 1  k  l+1  j\rangle \\&&+  x_2  z_1 \langle 1  k  l+1  j+1\rangle + 
  w_2  y_1 \langle 1  k+1  l+1  j\rangle +  w_2  z_1 \langle 1  k+1  l+1  j+1\rangle +
   x_1  x_2  y_1 \langle i  k  l+1  j\rangle \\&&+  x_1  x_2  z_1 \langle i  k  l+1  j+1\rangle + 
  w_2  x_1  y_1 \langle i  k+1  l+1  j\rangle + 
  w_2  x_1  z_1 \langle i  k+1  l+1  j+1\rangle + 
  w_1  x_2  y_1 \langle i+1  k  l+1  j\rangle \\&&+ 
  w_1  x_2  z_1 \langle i+1  k  l+1  j+1\rangle + 
  w_1  w_2  y_1 \langle i+1  k+1  l+1  j\rangle + 
  w_1  w_2  z_1 \langle i+1  k+1  l+1  j+1\rangle\\
  b&=&x_2 y_1  \langle 1  i+1  k  j \rangle  + x_2 z_1  \langle 1  i+1  k  j+1 \rangle  + 
 y_1 y_2  \langle 1  i+1  l  j \rangle  + y_2 z_1  \langle 1  i+1  l  j+1 \rangle  + 
 y_1 z_2  \langle 1  i+1  l+1  j \rangle  \\&&+ z_1 z_2  \langle 1  i+1  l+1  j+1 \rangle \\
 c&=&x_2 y_1 \langle 1  i  k  j\rangle  + x_2 z_1 \langle 1  i  k  j+1\rangle  + 
 w_2 y_1 \langle 1  i  k+1  j\rangle  + w_2 z_1 \langle 1  i  k+1  j+1\rangle  + 
 y_1 y_2 \langle 1  i  l  j\rangle  + y_2 z_1 \langle 1  i  l +1  j\rangle \\&& + 
 y_1 z_2 \langle 1  i  l+1  j\rangle  + z_1 z_2 \langle 1  i  l+1  j+1\rangle \\
 d&=&w_1 y_1 \langle 1  i+1  k+1  j\rangle  + w_1 z_1 \langle 1  i+1  k+1  j+1\rangle\\
 e&=& x_1  x_2  \langle 1  i  k  l \rangle  +  w_2  x_1  \langle 1  i  k+1  l \rangle  + 
  w_1  x_2  \langle 1  i+1  k  l \rangle  +  w_1  w_2  \langle 1  i+1  k+1  l \rangle  + 
  x_2  y_1  \langle 1  k  l  j \rangle  \\&&+  x_2  z_1  \langle 1  k  l  j+1 \rangle  + 
  w_2  y_1  \langle 1  k+1  l  j \rangle  +  w_2  z_1  \langle 1  k+1  l  j+1 \rangle  + 
  x_1  x_2  y_1  \langle i  k  l  j \rangle  +  x_1  x_2  z_1  \langle i  k  l  j+1 \rangle  \\&&+ 
  w_2  x_1  y_1  \langle i  k+1  l  j \rangle  +  w_2  x_1  z_1  \langle i  k+1  l  j+1 \rangle  + 
  w_1  x_2  y_1  \langle i+1  k  l  j \rangle  +  w_1  x_2  z_1  \langle i+1  k  l  j+1 \rangle  \\&&+ 
  w_1  w_2  y_1  \langle i+1  k+1  l  j \rangle  + 
  w_1  w_2  z_1  \langle i+1  k+1  l  j+1 \rangle  
\end{eqnarray*}and $a,b,c,d,e>0$. From $\langle ABCD \rangle>0$, 
\begin{eqnarray*}
z_2>\frac{b}{a}w_1+\frac{cx_1+dw_2-ey_2}{a}
\end{eqnarray*}
In the case of $cx_1+dw_2-ey_2>0$, 
\begin{eqnarray*}
z_2>\frac{b}{a}w_1+\frac{cx_1+dw_2-ey_2}{a},\ \ \text{and}\ \ y_2<\frac{cx_1+dw_2}{e} 
\end{eqnarray*}
Then the form of this sign pattern is
\begin{eqnarray*}
\Omega_1=\frac{1}{x_1}\frac{1}{x_2}\frac{1}{w_1}\frac{1}{w_2}\frac{1}{y_1}(\frac{1}{y_2}-\frac{1}{y_2-\frac{cx_1+dw_2}{e}})\frac{1}{z_1}\frac{1}{z_2-(\frac{b}{a}w_1+\frac{cx_1+dw_2-ey_2}{a} )}
\end{eqnarray*}
Another pattern is that $cx_1+dw_2-ey_2<0$,
\begin{eqnarray*}
w_1<\frac{a}{b}z_2-\frac{cx_1+dw_2-ey_2}{b},\ \ \text{and}\ \ y_2>\frac{cx_1+dw_2}{e}
\end{eqnarray*}
Then the form is
\begin{eqnarray*}
\Omega_2=\frac{1}{x_1}\frac{1}{x_2}(\frac{1}{w_1}-\frac{1}{w_1-(\frac{a}{b}z_2-\frac{cx_1+dw_2-ey_2}{b})})\frac{1}{w_2}\frac{1}{y_1}\frac{1}{y_2-\frac{cx_1+dw_2}{e}}\frac{1}{z_1}\frac{1}{z_2}
\end{eqnarray*}
The canonical form for this sign flip pattern is
\begin{equation}
(\Omega_1+\Omega_2)dx_1dx_2\cdots dz_1dz_2=\frac{dx_1dx_2\cdots dz_1dz_2}{x_1x_2w_1w_2y_1y_2z_1z_2}\frac{1}{(az_2-bw_1-cx_1-dw_2+ey_2)}\times\omega^1_{ijkl}
\end{equation}
\begin{eqnarray}
\omega_{ijkl}^1&=&\langle1   i  k  l\rangle  x_1  x_2  y_2 + \langle1   i  k  l + 1\rangle  x_1  x_2  z_2 + 
 \langle1   i  k + 1  l\rangle  w_2  x_1  y_2 + \langle1   i  k + 1  l + 1\rangle  w_2  x_1  z_2\nonumber \\
 &+& 
 \langle1   i + 1  k  l\rangle  w_1  x_2  y_2 + \langle1   i + 1  k  l + 1\rangle  w_1  x_2  z_2 + 
 \langle1   i + 1  k + 1  l\rangle  w_1  w_2  y_2 + 
 \langle1   i + 1  k + 1  l + 1\rangle  w_1  w_2  z_2 \nonumber\\
 &+& \langle1  k  l  j\rangle  x_2  y_1  y_2 + 
 \langle1  k  l  j + 1\rangle  x_2  y_2  z_1 + \langle1  k  l + 1  j\rangle  x_2  y_1  z_2 + 
 \langle1  k  l + 1  j + 1\rangle  x_2  z_1  z_2\nonumber\\
 & +& \langle1  k + 1  l  j\rangle  w_2  y_1  y_2 + 
 \langle1  k + 1  l  j + 1\rangle  w_2  y_2  z_1 + \langle1  k + 1  l + 1  j\rangle  w_2  y_1  z_2 \nonumber\\
 &+& 
 \langle1  k + 1  l + 1  j + 1\rangle  w_2  z_1  z_2 + \langle i  k  l  j\rangle  x_1  x_2  y_1  y_2 + 
 \langle i  k  l  j + 1\rangle  x_1  x_2  y_2  z_1 + \langle i  k  l + 1  j\rangle  x_1  x_2  y_1  z_2\nonumber \\
 &+& 
 \langle i  k  l + 1  j + 1\rangle  x_1  x_2  z_1  z_2 + \langle i  k + 1  l  j\rangle  w_2  x_1  y_1  y_2 +
  \langle i  k + 1  l  j + 1\rangle  w_2  x_1  z_1  y_2\nonumber \\&+& 
 \langle i  k + 1  l + 1  j\rangle  w_2  x_1  y_1  z_2 + 
 \langle i  k + 1  l + 1  j + 1\rangle  w_2  x_1  z_1  z_2 + 
 \langle i + 1  k + 1  l  j\rangle  w_2  w_1  y_1  y_2\nonumber \\&+& 
 \langle i + 1  k + 1  l  j + 1\rangle  w_2  w_1  z_1  y_2 + 
 \langle i + 1  k + 1  l + 1  j\rangle  w_2  w_1  y_1  z_2 + 
 \langle i + 1  k + 1  l + 1  j + 1\rangle  w_2  w_1  z_1  z_2\nonumber \\&+& 
 \langle i + 1  k  l  j\rangle  w_1  x_2  y_1  y_2 + \langle i + 1  k  l  j + 1\rangle  w_1  x_2  y_2  z_1 +
  \langle i + 1  k  l + 1  j\rangle  w_1  x_2  y_1  z_2\nonumber \\&+& 
 \langle i + 1  k  l + 1  j + 1\rangle  w_1  x_2  z_1  z_2
 \end{eqnarray}
 In the momentum twistor space, 
 \begin{equation}
\Omega^1_{ijkl}=\frac{\langle 1ii+1 A_j \rangle\langle1kk+1 C_l \rangle\langle AB d^2A \rangle\langle AB d^2B \rangle\langle CD d^2C \rangle\langle CD d^2D \rangle}{\langle AB 1i \rangle\langle AB 1i+1 \rangle\langle AB 1j \rangle\langle AB 1j+1 \rangle\langle ABCD \rangle\langle CD 1k \rangle\langle CD 1k+1 \rangle\langle CD1l \rangle\langle CD 1l+1 \rangle}\times\omega^{1'}_{ijkl}.
\end{equation}
for
  \begin{equation}
\omega_{ijkl}^{1'}=\frac{\langle ABii+1 \rangle\langle A_j C_k C_l 1\rangle+\langle A_iA_jC_k C_l  \rangle}{\langle ABii+1\rangle\langle ABjj+1\rangle\langle CDkk+1\rangle\langle CDll+1\rangle}.
\end{equation}
 Another forms can be obtained similarly
  \begin{eqnarray}
\omega_{ijkl}^2&=& \langle 1   i   k   l \rangle   x_1  x_2  y_2 +  \langle 1   i   k   l + 1 \rangle   x_1  x_2  z_2 + 
  \langle 1   i   k + 1   l \rangle   w_2  x_1  y_2 +  \langle 1   i   k + 1   l + 1 \rangle   w_2  x_1  z_2 + 
  \langle 1   i   j   l \rangle   x_1  y_1  y_2 \nonumber\\
  &+&  \langle 1   i   j   l + 1 \rangle   x_1  y_1  z_2 + 
  \langle 1   i   j + 1   l \rangle   z_1  x_1  y_2 +  \langle 1   i   j + 1   l + 1 \rangle   z_1  x_1  z_2 + 
  \langle 1   i + 1   k   l \rangle   w_1  x_2  y_2 \nonumber\\
  &+&  \langle 1   i + 1   k   l + 1 \rangle   w_1  x_2  z_2 + 
  \langle 1   i + 1   k + 1   l \rangle   w_2  w_1  y_2 + 
  \langle 1   i + 1   k + 1   l + 1 \rangle   w_2  w_1  z_2 +  \langle 1   i + 1   j   l \rangle   w_1  y_1  y_2 \nonumber\\
  &+& 
  \langle 1   i + 1   j   l + 1 \rangle   w_1  y_1  z_2 +  \langle 1   i + 1   j + 1   l \rangle   z_1  w_1  y_2 + 
  \langle 1   i + 1   j + 1   l + 1 \rangle   z_1  w_1  z_2
 \end{eqnarray}
  \begin{eqnarray}
\omega_{ijkl}^3&=& \langle 1   i   j   k\rangle   x_1  x_2  y_1 + \langle 1   i   j   k + 1\rangle   w_2  x_1  y_1 + 
 \langle 1   i   j   l\rangle   x_1  y_1  y_2 + \langle 1   i   j   l + 1\rangle   x_1  y_1  z_2 + 
 \langle 1   i   j + 1   k\rangle   x_1  x_2  z_1 \nonumber\\&+& \langle 1   i   j + 1   k + 1\rangle   w_2  x_1  z_1 + 
 \langle 1   i   j + 1   l\rangle   x_1  y_2  z_1 + \langle 1   i   j + 1   l + 1\rangle   x_1  z_1  z_2 + 
 \langle 1   i   k   l\rangle   x_1  x_2  y_2 \nonumber\\&+& \langle 1   i   k   l + 1\rangle   x_1  x_2  z_2 + 
 \langle 1   i   k + 1   l\rangle   w_2  x_1  y_2 + \langle 1   i   k + 1   l + 1\rangle   w_2  x_1  z_2 + 
 \langle 1   i + 1   j   k\rangle   w_1  x_2  y_1 \nonumber\\&+& \langle 1   i + 1   j   k + 1\rangle   w_2  w_1  y_1 + 
 \langle 1   i + 1   j   l\rangle   w_1  y_1  y_2 + \langle 1   i + 1   j   l + 1\rangle   w_1  y_1  z_2 + 
 \langle 1   i + 1   j + 1   k\rangle   w_1  x_2  z_1 \nonumber\\&+& 
 \langle 1   i + 1   j + 1   k + 1\rangle   w_2  w_1  z_1 + 
 \langle 1   i + 1   j + 1   l\rangle   w_1  y_2  z_1 + 
 \langle 1   i + 1   j + 1   l + 1\rangle   w_1  z_1  z_2 \nonumber\\&+& \langle 1   i + 1   k   l\rangle   w_1  x_2  y_2 + 
 \langle 1   i + 1   k   l + 1\rangle   w_1  x_2  z_2 + \langle 1   i + 1   k + 1   l\rangle   w_2  w_1  y_2 + 
 \langle 1   i + 1   k + 1   l + 1\rangle   w_2  w_1  z_2 \nonumber\\&+& \langle 1   j   k   l\rangle   x_2  y_1  y_2 + 
 \langle 1   j   k   l + 1\rangle   x_2  y_1  z_2 + \langle 1   j   k + 1   l\rangle   w_2  y_1  y_2 + 
 \langle 1   j   k + 1   l + 1\rangle   w_2  y_1  z_2 + \langle 1   j + 1   k   l\rangle   x_2  z_1  y_2 \nonumber\\&+& 
 \langle 1   j + 1   k   l + 1\rangle   x_2  z_1  z_2 + \langle 1   j + 1   k + 1   l\rangle   w_2  z_1  y_2 + 
 \langle 1   j + 1   k + 1   l + 1\rangle   w_2  z_1  z_2\nonumber\\& +& \langle i   j   k   l\rangle   x_1  x_2  y_1  y_2 + 
 \langle i   j   k   l + 1\rangle   x_1  x_2  y_1  z_2 + \langle i   j   k + 1   l\rangle   w_2  x_1  y_1  y_2 + 
 \langle i   j   k + 1   l + 1\rangle   w_2  x_1  y_1  z_2 \nonumber\\&+& \langle i   j + 1   k   l\rangle   x_1  x_2  z_1  y_2 +
  \langle i   j + 1   k   l + 1\rangle   x_1  x_2  z_1  z_2 + 
 \langle i   j + 1   k + 1   l\rangle   w_2  x_1  z_1  y_2 \nonumber\\&+& 
 \langle i   j + 1   k + 1   l + 1\rangle   w_2  x_1  z_1  z_2 + 
 \langle i + 1   j   k   l\rangle   w_1  x_2  y_1  y_2 + \langle i + 1   j   k   l + 1\rangle   w_1  x_2  y_1  z_2\nonumber \\&+&
  \langle i + 1   j   k + 1   l\rangle   w_2  w_1  y_1  y_2 + 
 \langle i + 1   j   k + 1   l + 1\rangle   w_2  w_1  y_1  z_2 + 
 \langle i + 1   j + 1   k   l\rangle   w_1  x_2  z_1  y_2 \\&+& 
 \langle i + 1   j + 1   k   l + 1\rangle   w_1  x_2  z_1  z_2 + 
 \langle i + 1   j + 1   k + 1   l\rangle   w_2  w_1  z_1  y_2 + 
 \langle i + 1   j + 1   k + 1   l + 1\rangle   w_2  w_1  z_1  z_2\nonumber
 \end{eqnarray}
   \begin{eqnarray}
\omega_{ijkl}^4&=&\langle 1  i  i + 1  l\rangle  w_2 x_1 y_2 + \langle 1  i  i + 1  l + 1\rangle  w_2 x_1 z_2 + 
 \langle 1  i  i + 1  j\rangle  w_1 x_2 y_1 + \langle 1  i  i + 1  j + 1\rangle  w_1 x_2 z_1\nonumber \\&+& 
 \langle 1  i  l  j\rangle  x_2 y_1 y_2 + \langle 1  i  l  j + 1\rangle  x_2 y_2 z_1 + 
 \langle 1  i  l + 1  j\rangle  x_2 y_1 z_2 + \langle 1  i  l + 1  j + 1\rangle  x_2 z_1 z_2 + 
 \langle 1  i + 1  l  j\rangle  w_2 y_1 y_2 \nonumber\\&+& \langle 1  i + 1  l  j + 1\rangle  w_2 y_2 z_1 + 
 \langle 1  i + 1  l + 1  j\rangle  w_2 y_1 z_2 + 
 \langle 1  i + 1  l + 1  j + 1\rangle  w_2 z_1 z_2 + 
 \langle i  i + 1  l  j\rangle  w_2 x_1 y_1 y_2 \nonumber\\&+& \langle i  i + 1  l  j + 1\rangle  w_2 x_1 y_2 z_1 +
  \langle i  i + 1  l + 1  j\rangle  w_2 x_1 y_1 z_2 + 
 \langle i  i + 1  l + 1  j + 1\rangle  w_2 x_1 z_1 z_2
 \end{eqnarray}
 \begin{eqnarray}
\omega_{ijkl}^5&=&\langle 1  i  i + 1  j\rangle   (w_1 x_2 y_1 + w_2 x_1 y_2) + 
 \langle 1  i  i + 1  j + 1\rangle   (w_1 x_2 z_1 + w_2 x_1 z_2) + 
 \langle 1  i  j  j + 1\rangle   (x_2 y_2 z_1 + x_1 y_1 z_2) \nonumber\\&+& 
 \langle 1  i + 1  j  j + 1\rangle   (w_2 y_2 z_1 + w_1 y_1 z_2) + 
 \langle i  i + 1  j  j + 1\rangle   (w_2 x_1 y_2 z_1 + w_1 x_2 y_1 z_2)
\end{eqnarray}
 \begin{eqnarray}
\omega_{ijkl}^6&=&\langle 1  i  k  j\rangle  x_1 x_2 y_2 + \langle 1  i  k  j + 1\rangle  x_1 x_2 z_2 + 
 \langle 1  i  k + 1  j\rangle  w_2 x_1 y_2 + \langle 1  i  k + 1  j + 1\rangle  w_2 x_1 z_2 + 
 \langle 1  i  j  j + 1\rangle  x_1 y_1 z_2\nonumber \\
 &+& \langle 1  i + 1  k  j\rangle  w_1 x_2 y_2 + 
 \langle 1  i + 1  k  j + 1\rangle  w_1 x_2 z_2 + \langle 1  i + 1  k + 1  j\rangle  w_2 w_1 y_2 + 
 \langle 1  i + 1  k + 1  j + 1\rangle  w_2 w_1 z_2 \nonumber\\&+& 
 \langle 1  i + 1  j  j + 1\rangle  w_1 y_1 z_2 + \langle 1  k  j  j + 1\rangle  x_2 y_2 z_1 + 
 \langle 1  k + 1  j  j + 1\rangle  w_2 y_2 z_1 + \langle i  k  j  j + 1\rangle  x_1 x_2 y_2 z_1 \nonumber\\&+& 
 \langle i  k + 1  j  j + 1\rangle  w_2 x_1 y_2 z_1 + 
 \langle i + 1  k  j  j + 1\rangle  w_1 x_2 y_2 z_1 + 
 \langle i + 1  k + 1  j  j + 1\rangle  w_1 w_2 y_2 z_1
\end{eqnarray}
 \begin{eqnarray}
\omega_{ijkl}^7&=&\langle 1  i  j  j + 1\rangle  w_2 x_1 y_1 + \langle 1  i  j  l\rangle  (x_1 x_2 y_2 + x_1 y_1 y_2) + 
 \langle 1  i  j  l + 1\rangle  (x_1 x_2 z_2 + x_1 y_1 z_2)\nonumber\\& +& 
 \langle 1  i  j + 1  l\rangle  (w_2 x_1 y_2 + x_1 y_2 z_1) + 
 \langle 1  i  j + 1  l + 1\rangle  (w_2 x_1 z_2 + x_1 z_1 z_2) + 
 \langle 1  i + 1  j  j + 1\rangle  w_2 w_1 y_1 \nonumber\\&+& 
 \langle 1  i + 1  j  l\rangle  (w_1 x_2 y_2 + w_1 y_1 y_2) + 
 \langle 1  i + 1  j  l + 1\rangle  (w_1 x_2 z_2 + w_1 y_1 z_2) \nonumber\\&+& 
 \langle 1  i + 1  j + 1  l\rangle  (w_2 w_1 y_2 + w_1 y_2 z_1) + 
 \langle 1  i + 1  j + 1  l + 1\rangle  (w_2 w_1 z_2 + w_1 z_1 z_2) + 
 \langle 1  j  j + 1  l\rangle  w_2 y_1 y_2 \nonumber\\&+& \langle 1  j  j + 1  l + 1\rangle  w_2 y_1 z_2 + 
 \langle i  j  j + 1  l\rangle  w_2 x_1 y_1 y_2 + \langle i  j  j + 1  l + 1\rangle  w_2 x_1 y_1 z_2 +
  \langle i + 1  j  j + 1  l\rangle  w_1 w_2 y_1 y_2 \nonumber\\&+& 
 \langle i + 1  j  j + 1  l + 1\rangle  w_1 w_2 y_1 z_2
\end{eqnarray}
 \begin{eqnarray}
\omega_{ijkl}^8=\omega_{ijkl}^4,&&\omega_{ijkl}^9=\omega_{ijkl}^7,\ \ \ \omega_{ijkl}^{10}=\omega_{ijkl}^2,\ \ \ \omega_{ijkl}^{11}=\omega_{ijkl}^6,\ \ \ \omega_{ijkl}^{12}=\omega_{ijkl}^1,\ \ \ \omega_{ijkl}^{13}=\omega_{ijkl}^3,\ \nonumber\\&&(x_1,w_1,y_1,z_1) \leftrightarrow (x_2,w_2,y_2,z_2) , (i,j) \leftrightarrow (k,l).
\end{eqnarray}
Next, we rewrite these forms in the momentum twistor space
  \begin{equation}
\omega_{ijkl}^{1'}=\frac{\langle ABii+1 \rangle\langle A_j C_k C_l 1\rangle+\langle A_iA_jC_k C_l  \rangle}{\langle ABii+1\rangle\langle ABjj+1\rangle\langle CDkk+1\rangle\langle CDll+1\rangle}.
\end{equation}
  \begin{equation}
\omega_{ijkl}^{2'}=\frac{-\langle ABjj+1 \rangle\langle  A_i C_k C_l 1\rangle +\langle CDkk+1 \rangle\langle A_i A_j C_l 1\rangle}{\langle ABii+1\rangle\langle ABjj+1\rangle\langle CDkk+1\rangle\langle CDll+1\rangle}
\end{equation}
 \begin{eqnarray}
\omega_{ijkl}^{3'}&=&\frac{1}{\langle ABii+1\rangle\langle ABjj+1\rangle\langle CDkk+1\rangle\langle CDll+1\rangle}\nonumber\\
&\times&\{\langle ABii+1 \rangle\langle  A_j C_k C_l 1\rangle -\langle ABjj+1 \rangle\langle  A_i C_k C_l 1\rangle-\langle AB 1i  \rangle\langle i+1 A_j C_k  C_l  \rangle\nonumber\\
&+&\langle CDkk+1 \rangle\langle  A_i A_j C_l 1\rangle -\langle CDll+1 \rangle\langle  A_i A_j C_k 1\rangle+\langle AB 1i+1  \rangle\langle i A_j C_k  C_l  \rangle\}\nonumber\\
\end{eqnarray}
 \begin{eqnarray}
\omega_{ijkl}^{4'}&=&\frac{1}{\langle ABii+1\rangle\langle ABjj+1\rangle\langle CDii+1\rangle\langle CDll+1\rangle}\nonumber\\
&\times&\{\langle ABii+1 \rangle\langle  A_j C_k C_l 1\rangle +\langle AB1i+1 \rangle\langle CD1i \rangle\langle  A_j C_l ii+1\rangle \\
&+&\langle AB1i \rangle\langle CD1i+1 \rangle\langle CDll+1 \rangle\langle A_j 1ii+1\rangle+\langle AB1i+1 \rangle\langle CD1i \rangle\langle ABjj+1 \rangle\langle C_l 1ii+1\rangle\}\nonumber
\end{eqnarray}
 \begin{eqnarray}
\omega_{ijkl}^{5'}&=&\frac{1}{\langle ABii+1\rangle\langle ABjj+1\rangle\langle CDii+1\rangle\langle CDjj+1\rangle}\nonumber\\
&\times&\{\langle 1ii+1 A_j\rangle(\langle AB1i \rangle\langle CD1k+1 \rangle\langle CDll+1 \rangle+\langle AB1j+1 \rangle\langle CD1l \rangle\langle CDkk+1 \rangle)\\
&+&\langle 1kk+1 C_l\rangle(\langle AB1i+1 \rangle\langle ABjj+1 \rangle\langle CD1k \rangle+\langle AB1j \rangle\langle ABii+1 \rangle\langle CD1l+1 \rangle)\nonumber\\
&+&\langle ii+1jj+1 \rangle(\langle AB1i \rangle\langle AB1j+1 \rangle\langle CD1k+1 \rangle\langle CD1l \rangle
+\langle AB1i+1 \rangle\langle AB1j \rangle\langle CD1k \rangle\langle CD1l+1 \rangle)\}\nonumber
\end{eqnarray}
\begin{eqnarray}
\omega_{ijkl}^{6'}&=&\frac{1}{\langle ABii+1\rangle\langle ABjj+1\rangle\langle CDkk+1\rangle\langle CDjj+1\rangle}\nonumber\\
&\times&\{\langle AB1j \rangle\langle ABii+1 \rangle\langle CD1l+1 \rangle\langle C_k 1jj+1\rangle+\langle AB1j+1 \rangle\langle CD1l \rangle\langle CDkk+1 \rangle\langle A_i 1jj+1\rangle\nonumber\\
&+&\langle ABjj+1 \rangle (\langle CD 1l+1 \rangle \langle A_i C_k 1j\rangle-\langle CD 1l \rangle \langle A_i C_k 1j+1\rangle)\}
\end{eqnarray}
\begin{eqnarray}
\omega_{ijkl}^{7'}&=&\frac{1}{\langle ABii+1\rangle\langle ABjj+1\rangle\langle CDjj+1\rangle\langle CDll+1\rangle}\nonumber\\
&\times&\{\langle AB1j \rangle\langle CD1 j+1\rangle\langle AB1i \rangle\langle i+1jj+1C_k \rangle -\langle AB1j \rangle\langle CD1 j+1\rangle\langle AB1i+1 \rangle\langle ijj+1C_k \rangle\nonumber\\
&+&\langle AB1j+1\rangle \langle CD 1j \rangle \langle CD kk+1 \rangle \langle 1jj+1A_i\rangle+\langle AB1j\rangle \langle CD 1j+1 \rangle \langle AB ii+1 \rangle \langle 1jj+1C_k\rangle\nonumber\\
&+&\langle ABjj+1 \rangle\langle 1 C_k C_j A_i \rangle\}
\end{eqnarray}
\begin{eqnarray}
\omega_{ijkl}^{8'}=\omega_{ijkl}^{4'},&&\omega_{ijkl}^{9'}=\omega_{ijkl}^{7'},\ \ \ \omega_{ijkl}^{10'}=\omega_{ijkl}^{2'},\ \ \ \omega_{ijkl}^{11'}=\omega_{ijkl}^{6'},\ \ \ \omega_{ijkl}^{12'}=\omega_{ijkl}^{1'},\ \ \ \omega_{ijkl}^{13'}=\omega_{ijkl}^{3'},\ \nonumber\\&& (AB) \leftrightarrow (CD), (i,j) \leftrightarrow (k,l)
\end{eqnarray}
\section{Explicit Results of the 2-loop n-point MHV Log Amplituhedron}
\begin{equation}
\Omega[\log{[\mathcal{A}^{n\text{-pt 2-loop}}_{\text{MHV}}]}]=\sum_{\substack{i,j,k,l=2,3,\cdots,n-1\\i<k<l<j}}\Omega_{ijkl}^1[\log]+\sum_{i<k<j<l}\Omega_{ijkl}^2[\log]\cdots+\sum_{k<l<i<j}\Omega_{ijkl}^{13}[\log]
\end{equation}
for
\begin{equation}
\Omega^{m}_{ijkl}[\log]=\frac{dx_1dx_2\cdots dz_1dz_2}{x_1x_2w_1w_2y_1y_2z_1z_2}\frac{-1}{(az_2-bw_1-cx_1-dw_2+ey_2)}\times\omega^m_{ijkl}[\log]
\end{equation}
where
\begin{eqnarray}
\omega^1_{ijkl}[\log]&=&x_2 y_1  \langle 1  i+1  k  j \rangle  + x_2 z_1  \langle 1  i+1  k  j+1 \rangle  + 
 y_1 y_2  \langle 1  i+1  l  j \rangle  + y_2 z_1  \langle 1  i+1  l  j+1 \rangle \nonumber \\&+& 
 y_1 z_2  \langle 1  i+1  l+1  j \rangle+ z_1 z_2  \langle 1  i+1  l+1  j+1 \rangle+ x_2 y_1 \langle 1  i  k  j\rangle  + x_2 z_1 \langle 1  i  k  j+1\rangle  \nonumber\\ &+& 
 w_2 y_1 \langle 1  i  k+1  j\rangle  + w_2 z_1 \langle 1  i  k+1  j+1\rangle  + 
 y_1 y_2 \langle 1  i  l  j\rangle  + y_2 z_1 \langle 1  i  l +1  j\rangle  + 
 y_1 z_2 \langle 1  i  l+1  j\rangle   \nonumber\\&+& z_1 z_2 \langle 1  i  l+1  j+1\rangle +w_1 y_1 \langle 1  i+1  k+1  j\rangle  + w_1 z_1 \langle 1  i+1  k+1  j+1\rangle
\end{eqnarray}
\begin{eqnarray}
\omega^2_{ijkl}[\log]&=&-x_1 x_2 y_1 \langle 1  i  k  j\rangle   - x_1 x_2 z_1 \langle 1  i  k  j + 1\rangle   - 
 w_2 x_1 y_1 \langle 1  i  k + 1  j\rangle \nonumber\\ &-& w_2 x_1 z_1 \langle 1  i  k + 1  j + 1\rangle   - 
 w_1 x_2 y_1 \langle 1  i + 1  k  j\rangle   - w_1 x_2 z_1 \langle 1  i + 1  k  j + 1\rangle   - 
 w_1 w_2 y_1 \langle 1  i + 1  k + 1  j\rangle   \nonumber\\&-& 
 w_1 w_2 z_1 \langle 1  i + 1  k + 1  j + 1\rangle   - x_2 y_1 y_2 \langle 1  k  j  l\rangle   - 
 x_2 y_1 z_2 \langle 1  k  j  l + 1\rangle   - x_2 y_2 z_1 \langle 1  k  j + 1  l\rangle   \nonumber\\&-& 
 x_2 z_1 z_2 \langle 1  k  j + 1  l + 1\rangle   - w_2 y_1 y_2 \langle 1  k + 1  j  l\rangle   - 
 w_2 y_1 z_2 \langle 1  k + 1  j  l + 1\rangle   - w_2 y_2 z_1 \langle 1  k + 1  j + 1  l\rangle   \nonumber\\&-& 
 w_2 z_1 z_2 \langle 1  k + 1  j + 1  l + 1\rangle   - x_1 x_2 y_1 y_2 \langle i  k  j  l\rangle   - 
 x_1 x_2 y_1 z_2 \langle i  k  j  l + 1\rangle   - x_1 x_2 y_2 z_1 \langle i  k  j + 1  l\rangle   \nonumber\\&-& 
 x_1 x_2 z_1 z_2 \langle i  k  j + 1  l + 1\rangle   - w_2 x_1 y_1 y_2 \langle i  k + 1  j  l\rangle   -
  w_2 x_1 y_1 z_2 \langle i  k + 1  j  l + 1\rangle  \nonumber\\&-& 
 w_2 x_1 y_2 z_1 \langle i  k + 1  j + 1  l\rangle   - 
 w_2 x_1 z_1 z_2 \langle i  k + 1  j + 1  l + 1\rangle   - 
 w_1 x_2 y_1 y_2 \langle i + 1  k  j  l\rangle  \nonumber \\&-& w_1 x_2 y_1 z_2 \langle i + 1  k  j  l + 1\rangle   -
  w_1 x_2 y_2 z_1 \langle i + 1  k  j + 1  l\rangle   - 
 w_1 x_2 z_1 z_2 \langle i + 1  k  j + 1  l + 1\rangle   \nonumber\\&-& 
 w_1 w_2 y_1 y_2 \langle i + 1  k + 1  j  l\rangle   - 
 w_1 w_2 y_1 z_2 \langle i + 1  k + 1  j  l + 1\rangle   - 
 w_1 w_2 y_2 z_1 \langle i + 1  k + 1  j + 1  l\rangle   \nonumber\\&-& 
 w_1 w_2 z_1 z_2 \langle i + 1  k + 1  j + 1  l + 1\rangle
 \end{eqnarray}
\begin{eqnarray}
\omega^3_{ijkl}[\log]&=&0
\end{eqnarray}
\begin{eqnarray}
\omega^4_{ijkl}[\log]&=&- w_1  x_2  y_2 \langle   1  i  i + 1  l\rangle   -  w_1  x_2  z_2 \langle   1  i  i + 1  l + 1\rangle   - 
  w_2  x_1  y_1 \langle   1  i  i + 1  j\rangle   -  w_2  x_1  z_1 \langle   1  i  i + 1  j + 1\rangle   \nonumber\\&-& 
  x_1  y_1  y_2 \langle   1  i  l  j\rangle   -  x_1  y_2  z_1 \langle   1  i  l  j + 1\rangle   - 
  x_1  y_1  z_2 \langle   1  i  l + 1  j\rangle   -  x_1  z_1  z_2 \langle   1  i  l + 1  j + 1\rangle   - 
  w_1  y_1  y_2 \langle   1  i + 1  l  j\rangle  \nonumber\\&-&  w_1  y_2  z_1 \langle   1  i + 1  l  j + 1\rangle   - 
  w_1  y_1  z_2 \langle   1  i + 1  l + 1  j\rangle   - 
  w_1  z_1  z_2 \langle   1  i + 1  l + 1  j + 1\rangle   - 
  w_1  x_2  y_1  y_2 \langle   i  i + 1  l  j\rangle   \nonumber\\&-&  w_1  x_2  y_2  z_1 \langle   i  i + 1  l  j + 1\rangle   -
   w_1  x_2  y_1  z_2 \langle   i  i + 1  l + 1  j\rangle   - 
  w_1  x_2  z_1  z_2 \langle   i  i + 1  l + 1  j + 1\rangle
  \end{eqnarray}
\begin{eqnarray}
  \omega^5_{ijkl}[\log]&=&- w_2  x_1  y_1  \langle 1   i   i + 1   j \rangle   -  w_1  x_2  y_2  \langle 1   i   i + 1   j \rangle   - 
  w_2  x_1  z_1  \langle 1   i   i + 1   j + 1 \rangle   -  w_1  x_2  z_2  \langle 1   i   i + 1   j + 1 \rangle   \nonumber\\&-& 
  x_1  y_2  z_1  \langle 1   i   j   j + 1 \rangle   -  x_2  y_1  z_2  \langle 1   i   j   j + 1 \rangle   - 
  w_1  y_2  z_1  \langle 1   i + 1   j   j + 1 \rangle   -  w_2  y_1  z_2  \langle 1   i + 1   j   j + 1 \rangle   \nonumber\\&-& 
  w_1  x_2  y_2  z_1  \langle i   i + 1   j   j + 1 \rangle   - 
  w_2  x_1  y_1  z_2  \langle i   i + 1   j   j + 1 \rangle
  \end{eqnarray}
\begin{eqnarray}
  \omega^6_{ijkl}[\log]&=&-w_2 x_1 y_1  \langle  1  i  i + 1  j \rangle   - w_2 x_1 z_1  \langle  1  i  i + 1  j + 1 \rangle   - 
 w_1 x_2 y_2  \langle  1  i  i + 1  l \rangle   - w_1 x_2 z_2  \langle  1  i  i + 1  l + 1 \rangle   \nonumber\\&-& 
 x_2 y_1 y_2  \langle  1  i  j  l \rangle   - x_2 y_1 z_2  \langle  1  i  j  l + 1 \rangle   - 
 x_2 y_2 z_1  \langle  1  i  j + 1  l \rangle   - x_2 z_1 z_2  \langle  1  i  j + 1  l + 1 \rangle   - 
 w_2 y_1 y_2  \langle  1  i + 1  j  l \rangle  \nonumber\\& -& w_2 y_1 z_2  \langle  1  i + 1  j  l + 1 \rangle   - 
 w_2 y_2 z_1  \langle  1  i + 1  j + 1  l \rangle   - 
 w_2 z_1 z_2  \langle  1  i + 1  j + 1  l + 1 \rangle   - 
 w_2 x_1 y_1 y_2  \langle  i  i + 1  j  l \rangle   \nonumber\\&-& w_2 x_1 y_1 z_2  \langle  i  i + 1  j  l + 1 \rangle   -
  w_2 x_1 y_2 z_1  \langle  i  i + 1  j + 1  l \rangle   - 
 w_2 x_1 z_1 z_2  \langle  i  i + 1  j + 1  l + 1 \rangle 
 \end{eqnarray}
\begin{eqnarray}
 \omega^7_{ijkl}[\log]&=&-x_1 x_2 y_1 \langle 1  i  k  j\rangle  - x_1 x_2 z_1 \langle 1  i  k  j + 1\rangle  - 
 w_2 x_1 y_1 \langle 1  i  k + 1  j\rangle  - w_2 x_1 z_1 \langle 1  i  k + 1  j + 1\rangle  \nonumber\\&-& 
 x_1 y_2 z_1 \langle 1  i  j  j + 1\rangle  - w_1 x_2 y_1 \langle 1  i + 1  k  j\rangle  - 
 w_1 x_2 z_1 \langle 1  i + 1  k  j + 1\rangle  - w_1 w_2 y_1 \langle 1  i + 1  k + 1  j\rangle  \nonumber\\&-& 
 w_1 w_2 z_1 \langle 1  i + 1  k + 1  j + 1\rangle  - 
 w_1 y_2 z_1 \langle 1  i + 1  j  j + 1\rangle  - x_2 y_1 z_2 \langle 1  k  j  j + 1\rangle  - 
 w_2 y_1 z_2 \langle 1  k + 1  j  j + 1\rangle  \nonumber\\&-& x_1 x_2 y_1 z_2 \langle i  k  j  j + 1\rangle  - 
 w_2 x_1 y_1 z_2 \langle i  k + 1  j  j + 1\rangle  - 
 w_1 x_2 y_1 z_2 \langle i + 1  k  j  j + 1\rangle  \nonumber\\&-& 
 w_1 w_2 y_1 z_2 \langle i + 1  k + 1  j  j + 1\rangle 
 \end{eqnarray}
  \begin{eqnarray}
\omega_{ijkl}^8[\log]=\omega_{ijkl}^4[\log],&&\omega_{ijkl}^9[\log]=\omega_{ijkl}^7[\log],\ \ \ \omega_{ijkl}^{10}[\log]=\omega_{ijkl}^2[\log],\ \ \ \omega_{ijkl}^{11}\log]=\omega_{ijkl}^6[\log],\nonumber\\&&\ \ \ \omega_{ijkl}^{12}[\log]=\omega_{ijkl}^1[\log],\ \ \ \omega_{ijkl}^{13}[\log]=0,\ \nonumber\\&&(x_1,w_1,y_1,z_1) \leftrightarrow (x_2,w_2,y_2,z_2) , (i,j) \leftrightarrow (k,l) 
\end{eqnarray}
We can similarly write these forms in the momentum twistor language.

\end{document}